\documentclass{article}

\usepackage{arxiv}

\usepackage[utf8]{inputenc} % allow utf-8 input
\usepackage[T1]{fontenc}    % use 8-bit T1 fonts
\usepackage{hyperref}       % hyperlinks
\usepackage{url}            % simple URL typesetting
\usepackage{booktabs}       % professional-quality tables
\usepackage{amsfonts}       % blackboard math symbols
\usepackage{nicefrac}       % compact symbols for 1/2, etc.
\usepackage{microtype}      % microtypography
\usepackage{lipsum}
\usepackage{graphicx}
\usepackage{subfig}
\usepackage{framed,multirow}

\graphicspath{ {./images/} }

\title{A multi--stage GAN for multi--organ chest X--ray image generation and segmentation}

\author{
 Giorgio Ciano \\
  Department of Information engineering \\
  University of Florence \\
  Florence, Italy \\
  \texttt{giorgio.ciano@unifi.it} \\
   \And
 Paolo Andreini \\
  Department of Information engineering and mathematics \\ 
  University of Siena \\ 
  Siena, Italy \\
  \texttt{paolo.andreini@unisi.it} \\
  \And
 Tommaso Mazzierli \\
  Department of Nephrology, AOU Careggi \\
  University of Florence \\ 
  Florence, Italy \\
  \texttt{tommaso.mazzierli@unifi.it} \\
  \And
  Monica Bianchini \\
  Department of Information engineering and mathematics \\ 
  University of Siena \\ 
  Siena, Italy \\
  \texttt{monica@diism.unisi.it} \\
  \And
  Franco Scarselli \\
  Department of Information engineering and mathematics \\ 
  University of Siena \\ 
  Siena, Italy \\
  \texttt{franco@diism.unisi.it} \\
}

\begin{document}
\maketitle
\begin{abstract}
Multi--organ segmentation of  X--ray images is of fundamental importance for computer aided diagnosis systems. However, the most advanced semantic segmentation methods rely on deep learning and require a huge amount of labeled images, which are rarely available due to both the high cost of human resources and the time required for labeling. In this paper, we present a novel multi--stage generation algorithm based on Generative Adversarial Networks (GANs) that can produce synthetic images along with their semantic labels and can be used for data augmentation. The main feature
of the method is that, unlike other approaches, generation occurs in several stages, which simplifies the procedure and allows it to be used on very small datasets.
The method has been evaluated on the segmentation of chest radiographic images, showing promising results. The multistage approach achieves state--of--the--art and, when very few images are used to train the GANs, outperforms the corresponding single--stage approach. 
\end{abstract}

% keywords can be removed

\section{Introduction}
\label{sec:introduction}
Chest X--ray (CXR) is one of the most used techniques worldwide for the diagnosis of various diseases, such as pneumonia, tuberculosis, infiltration, heart failure and lung cancer. Chest X--rays have enormous advantages: they are cheap,
X--ray equipment is also available in the poorest areas of the world and, moreover, the interpretation/reporting of X--rays is less operator--dependent than the results of other more advanced techniques, such as computed tomography (CT) and magnetic resonance (RMI).
 Furthermore, undergoing this examination is very fast and minimally invasive \cite{mettler2008effective}. Recently, CXR images have gained even greater importance due to COVID--19, which mainly causes lung infection and, after healing, often leaves widespread signs of pulmonary fibrosis: the respiratory tissue affected by the infection loses its characteristics and its normal structure. Consequently, CXR images are often used for the COVID--19 diagnosis and for the treatment of after-effects of SARS-CoV-2 \cite{hussain2021corodet, ismael2021deep, nayak2021application}.

Therefore, with the rapid growth in the number of CXRs performed per patient, there is an ever--increasing need for computer--aided diagnosis (CAD) systems to assist radiologists, since manual classification and annotation is time--consuming and subject to errors. Recently, deep learning (DL) has radically changed the perspective in medical image processing, and deep neural networks (DNNs) have been applied to a variety of tasks, including organ segmentation, object and lesion classification \cite{Ciano1}, image generation and registration \cite{van2006segmentation}. These DL methods constitute an important step towards the construction of CADs for medical images and, in particular, for CXRs. 

Semantic segmentation of anatomical structures is the process of classifying each pixel of an image according to the structure to which it belongs. In CAD, segmentation plays a fundamental role.
Indeed, segmentation of CXR images is usually
necessary to obtain regions of interest and allows to extract size measurements of organs (e.g., cardiothoracic ratio quantification) and irregular shapes, which can provide meaningful information on important diseases, such as cardiomegaly, emphysema and lung nodules \cite{qin2018computer}.
Segmentation may also help to improve the performance of automatic classification: in \cite{teixeira2020impact}, it is shown that, by exploiting segmentation, DL models focus their attention primarily on the lung, not taking into account unnecessary background information and noise. 

Modern state--of--the--art segmentation algorithms are largely based on DNNs \cite{long2015fully, chen2017deeplab, zhao2017pyramid}.
However, to achieve good results, DNNs need a fairly large amount of labeled data. Therefore, the main problem with segmentation by DNNs is the scarce availability of appropriate datasets to help solve a given task. This problem is even more evident in the medical field, where data availability is affected by privacy concerns and where a great deal of time and human resources are required to manually label each pixel of each image.

A common solution to cope with this problem is the generation of synthetic images, along with their semantic label--maps. This task can be carried out by Generative Adversarial Networks (GANs) \cite{goodfellow2014generative}, which can learn, using few training examples, the data distribution in a given domain. In this paper, we present  a new model, based on GANs, to generate multi--organ segmentation of CXR images. Unlike other approaches, the main feature of the proposed method is that generation occurs in three stages. In the first stage, the position of each anatomical part is generated
and represented by a``dot'' within the image; in the second stage, semantic labels are obtained from the dots; finally, the chest X--ray image is generated.
Each step is implemented by a GAN. More precisely, we adopt
Progressively Growing GANs (PGGANs) \cite{karras2017progressive}, a recent extension of GANs that allows 
to generate high resolution images, and pix2pixHD \cite{wang2018high} for the translation steps.
The intuitive idea underlying the approach is that generation benefits by the multi--stage procedure, since the GAN used in each single step faces a sub--problem, and can be trained  using fewer data.

In order to evaluate the performance of the proposed method,  synthetic images have been used to train a segmentation network\footnote{Here, we use the Segmentation Multiscale Attention Network (SMANet) \cite{bonechi2020weak}, a deep convolutional neural network based on the Pyramid Scene Parsing Network \cite{zhao2017pyramid}.}, subsequently applied on a popular benchmark for multi--organ chest segmentation, the Segmentation in Chest Radiographs (SCR) dataset \cite{van2006segmentation}. The results obtained are very promising and exceed (to the best of our knowledge) those obtained by other previous methods. Moreover, the quality of the produced segmentation has been confirmed by physicians. Finally, to demonstrate the capabilities of our approach, especially having little data available, we compared it to two other methods, using only $10\%$ of the images in the dataset. In particular, the multi--stage approach has been compared with a single--stage method --- in which chest X--ray images and semantic label--maps are generated simultaneously --- and with a two--stage
method --- where semantic label--maps are generated and then translated into X--ray images. 
The experimental results show that the proposed three--stage method outperforms the two--stage method, while 
the two--stage overcomes the single--stage approach, confirming that splitting the generation procedure can be advantageous, particularly when few training images are available.

The paper is organized as follows. In Section \ref{related}, the related literature is reviewed. Section \ref{generation} presents a description of the  proposed image generation method. Section \ref{experiments} shows and discusses the experimental results. Finally, in Section \ref{conclusions}, we draw some conclusions and describe future research.

\section{Related works}
 \label{related}
In the following, recent works related to the topics addressed in this paper are briefly reviewed, namely regarding synthetic image generation, image--to--image translation, and segmentation of medical images.
\subsection{Synthetic Image Generation}
Methods for generating images are by no means new and can be classified into two main categories: model--based and learning--based approaches. A model--based method consists in formulating a model of the observed data to render the image by a dedicated engine.
This approach has been widely adopted to generate images in many different domains \cite{andreini2019two, ANDREINI2020105268, 10.1007/978-3-030-01424-7_51}.
Nonetheless, the design of specialized engines for data generation requires a deep knowledge of the specific domain. For this reason, in recent years, the learning–based approach attracted increasing research interest. In this context, machine learning techniques are used to capture the intrinsic variability of a set of training images, so that the specific domain model is acquired implicitly from the data. Once the probability distribution that underlies the set of real images has been learned, the system can be used to generate new images that are likely to mimic the original ones. One of the most successful machine learning model for data generation is the Generative Adversarial Network (GAN) \cite{goodfellow2014generative}. A GAN is composed by two networks: a generator $G$ and a discriminator $D$. The former learns to generate data starting from a latent random variable $ \textbf{z} \in \mathbb{R}^{Z} $, while the latter aims at distinguishing real data from generated ones.
Training GANs is difficult, because it consists in a min--max game between two neural networks and convergence is not guaranteed. This problem is compounded in the generation of high resolution images, because the high resolution makes it easier to distinguish generated images from training images \cite{odena2017conditional}. One of the most successful approach to face this problem is represented by Progressively Growing GANs (PGGANs) \cite{karras2017progressive}. This model, in fact, is based on a multi--stage approach that aims to simplify and stabilize the training and allows to generate high resolution images. More specifically, in a PGGAN, the training starts at low resolution, while new blocks are progressively introduced in the system to increase the resolution of the generation. The generator and discriminator grow symmetrically until the desired resolution is reached. Based upon PGGANs, many different approaches have been proposed. For instance,  StyleGANs \cite{karras2019style} maintain the same discriminator as PGGANs, but introduce a new generator which is able to control the style of the generated images at different levels of detail. In StyleGAN2s \cite{karras2020analyzing}, an improved training scheme is introduced, which achieves the same goal --- training starts by focusing on low resolution images and then progressively shifts the focus to higher and higher resolutions --- without changing the network topology during training. In this way, the updated model shows improved results at the expense of longer training times and more computing resources.

In this paper, we use PGGANs in three different ways. For the single--stage method, a PGGAN simultaneously generates semantic label--maps and CXR images. For the two--stage method, only semantic label--maps are generated, while for the three--stage method we use a PGGAN to generate ``dots'' that correspond to different anatomical parts.

\subsection{Image--to--Image Translation}
Recently, beside image generation, adversarial learning has been also employed for image--to--image translation, whose goal is to translate an input image from one domain to another.
Many computer vision tasks, such as image super--resolution \cite{ledig2017photo}, image inpainting \cite{pathak2016context}, and style transfer \cite{gatys2015neural} can be casted into the image--to--image translation framework.
Both unsupervised \cite{liu2017unsupervised, liu2016coupled, yi2017dualgan, zhu2017unpaired} and supervised approaches \cite{isola2017image, karras2017progressive, chen2017photographic} can be used but, for the proposed application to CXR image generation, the unsupervised category is not relevant.
Supervised training uses a set of pairs of corresponding images $\left \{(s_i,t_i)\right \}$, where $s_i$ is an image of the source domain and $t_i$ is the corresponding image in the target domain. In the original GAN framework, there is no explicit way of controlling what to generate, since the output depends only on the latent vector $\textbf{z}$.
For this reason, in conditional GANs (cGANs) \cite{mirza2014conditional}, an additional input $\textbf{c}$ is introduced to guide the generation. In a cGAN, the generator can be  defined accordingly as $G(\textbf{c}, \textbf{z})$. Pix2Pix \cite{isola2017image} is a general approach for image--to--image translation and consists of a conditional GAN that operates in a supervised way. Pix2Pix uses a loss function that allows to generate plausible images in relation to the destination domain, which are also credible translations of the input image. With respect to supervised image--to--image translation techniques, in addition to the aforementioned Pix2Pix, the most used models are CRN \cite{chen2017photographic}, Pix2PixHD \cite{wang2018high}, BycicleGAN \cite{zhu2017toward}, SIMS \cite{qi2018semi}, and SPADE \cite{park2019semantic}.
In particular, Pix2PixHD \cite{wang2018high} improves upon Pix2Pix by employing a coarse--to--fine generator and discriminator, along with a feature--matching loss--function, allowing to translate images with higher resolution and quality. 

For the image--to--image translation phase, we use the Pix2PixHD network. The single--stage method does not require a translation step, while for the two--stage method we use Pix2PixHD to obtain a CXR image from the label--map. Finally, in the three--stage method, Pix2PixHD is used in two steps: for the translation from ``dots'' to semantic label--maps and, after, for the translation of label--maps into CXR images.

\subsection{Medical Image Generation}
In recent years, GANs have attracted the attention of medical researchers, their applications ranging from object detection \cite{sun2018adversarial, chen2018unsupervised, schlegl2017unsupervised} to registration \cite{zhang2020deformganan, FanCXYS18, tanner2018generative}, classification \cite{yi2018unsupervised, madani2018semi, lecouat2018semisupervised} and segmentation \cite{li2018cc, xue2018segan} of images. For instance, in \cite{frid2018gan}, different GANs have been used for the synthesis of each class of liver lesions (cysts, metastases and hemangiomas). However, in the medical domain, the use of complex machine learning models is often limited by the difficulty of collecting large sets of data. In this context, GANs can be employed to generate synthetic data, realizing a form of data augmentation. In fact, the GAN generated data can be used to enlarge the available datasets and to improve the performance in different tasks. 
As an example, GAN generated images have been successfully used to improve the performance in classification problems, by combining real and synthetic images during the training of a classifier.  
In \cite{hu2018unsupervised}, Wasserstein GANs (WGANs) and InfoGANs have been combined to classify histopathological images, whereas in \cite{yi2018unsupervised} WGAN and CatGAN generated images were used to improve the classification of dermoscopic images. Only in a few cases have GANs been used to generate chest radiographic images, as in \cite{madani2018semi}, where images for cardiac abnormality classification were obtained with a semi--supervised architecture, or  in \cite{srivastav2021improved}, where GANs were used to generate low resolution ($64\times{64}$) CXRs to diagnose Pneumonia. More related to this work, in \cite{andreini2019two}, high--resolution synthetic images of the retina and the corresponding semantic label--maps have been generated. Moreover, synthesizing images has been proven to be an effective method for data augmentation, that can be used to improve performance in retinal vessel segmentation. 

In this paper, chest X--ray images have been generated with the corresponding semantic label--maps (which correspond to different organs). We then used such images to train a segmentation network, with very promising results.

\subsection{Organ Segmentation}
X--rays are one of the most used techniques in medical diagnostics. The reasons are medical and economic, since they are cheap, non--invasive and fast examinations.
Many diseases, such as pneumonia, tuberculosis, lung cancer, and heart failure are commonly diagnosed from CXR images.  However, due to overlapping organs, low resolution and subtle anatomical shape and size variations, interpreting CXRs accurately remains challenging and requires highly qualified and trained personnel. Therefore, it is of a great clinical and scientific interest to develop computer--based systems that support the analysis of CXRs. In \cite{candemir2013lung}, a lung boundary detection system was proposed, building an anatomical atlas to be used in combination with graph--cut based image region refinement \cite{boykov2006graph, candemir2011statistical, boykovinteractive}.
A method for lung field segmentation, based on joint shape and appearance sparse learning, was proposed in \cite{shao2014hierarchical}, while  a technique for landmark detection was presented in \cite{ibragimov2016accurate}. Haar--like features and a random forest classifier were combined for the appearance of landmarks. Instead, a Gaussian distribution augmented by shape--based random forest classifiers was adopted for learning  spatial relationships between landmarks. \emph{InvertedNet}, an architecture able to segment the heart, clavicles and lungs, was introduced in \cite{novikov2018fully}. This network employs a loss function based on the Dice Coefficient, Exponential Linear Units (ELUs) activation functions, and a model architecture that aims at containing the number of parameters.
Moreover, the UNet \cite{ronneberger2015u} architecture has been widely used for lung segmentation, as in \cite{wang2017segmentation, oliveira2018deep, islam2018towards}. In the Structure Correcting Adversarial Network (SCAN) \cite{dai2018scan}  a segmentation network and a critic network were jointly trained  with an adversarial mechanism for organ segmentation in chest X--rays. 

\section{Chest X--Ray Generation} \label{generation}
The main goal of this study is to prove that by dividing the generation problem into multiple simpler stages, the quality of the generated images improves, so that they can be more effectively employed as a form of data augmentation. More specifically, we compare three different generation approaches. The first method, described in Section \ref{generation:single_step}, consists in generating chest X--ray images and the corresponding label--maps in a single stage. In the second approach, presented in Section \ref{generation:two_step}, the generation procedure is divided into two stages, where the label--maps are initially generated and then translated into images. The third method, reported in Section \ref{generation:three_step}, consists in a three stage approach, that starts by generating the position of the objects in the image, then the label--maps and, finally, the X--ray images. The images generated employing each of the three approaches are comparatively evaluated by training a segmentation network.

To increase the descriptive power of real images, especially with regards to the position of the various organs, standard data augmentation has preventively been applied. Therefore, the original X--ray images, along with their corresponding masks, were augmented by applying random rotations in the interval $[-2, 2]$ degrees, random horizontal, vertical and combined  translations from $-3\%$ to $+3\%$ of the number of pixels, and adding a Gaussian noise --- only to the original images ---, with $0$ mean and variance between $0.01$ and $0.03\times 255$.  For the generation of images, we essentially used two networks well known in the literature, namely PGGANs \cite{karras2017progressive} and Pix2PixHD \cite{wang2018high}, whose details are given in the following. In particular, in Sections \ref{generation:single_step}, \ref{generation:two_step}, and \ref{generation:three_step}, we extensively describe the three different generation procedures, respectively the single--stage, two--stage and three--stage methods. The next Section \ref{SMANET} presents the semantic segmentation network that has been employed. Finally, some details on the training method are collected in Section \ref{training}.

\subsection{Single--stage method} \label{generation:single_step}
\begin{figure}[!ht]
\begin{center}
\includegraphics[scale=0.35]{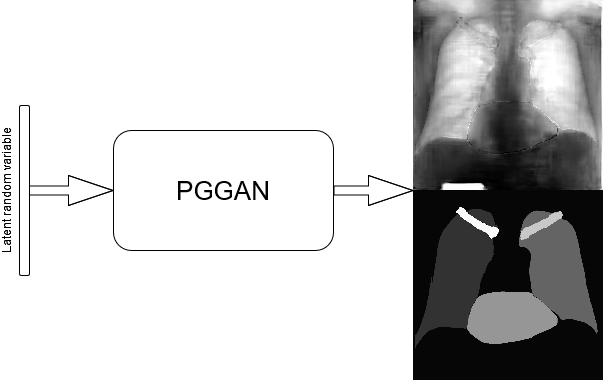}
\caption{The one--stage image generation scheme. The input of the network is a latent vector, while the PGGAN simultaneously produces the label--map and the X--ray image.}
\label{single_step_scheme}
\end{center}
\end{figure}

This baseline approach consists in stacking X--ray images and labels into two different channels, which are simultaneously fed into the PGGAN.
Therefore, the PGGAN is trained to generate pairs composed by an X--ray image and its corresponding label (see Figure \ref{single_step_scheme}).

\subsection{Two--stage method} \label{generation:two_step}
\begin{figure}[!ht]
\begin{center}
\includegraphics[scale=0.4]{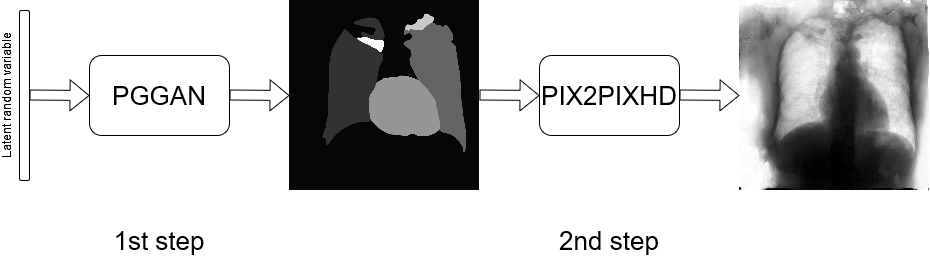}
\caption{The two--stage image generation scheme. In the first step, the PGGAN takes in input a latent vector and produces the label--map. The generated label--map is then used as input to a pix2pixHD module, which is trained to output the X--ray image.}
\label{two_step_scheme}
\end{center}
\end{figure}

In this approach, the generation procedure is divided into two steps. The first one consists in generating the labels through a PGGAN, while, in the second, the translation from the label to the corresponding chest X--ray image is carried out, using Pix2PixHD (see Figure \ref{two_step_scheme}).
\subsection{Three--stage method} \label{generation:three_step}
\begin{figure*}[!hb]

\vskip 0.2in
\begin{center}
\centerline{\includegraphics[scale=0.31]{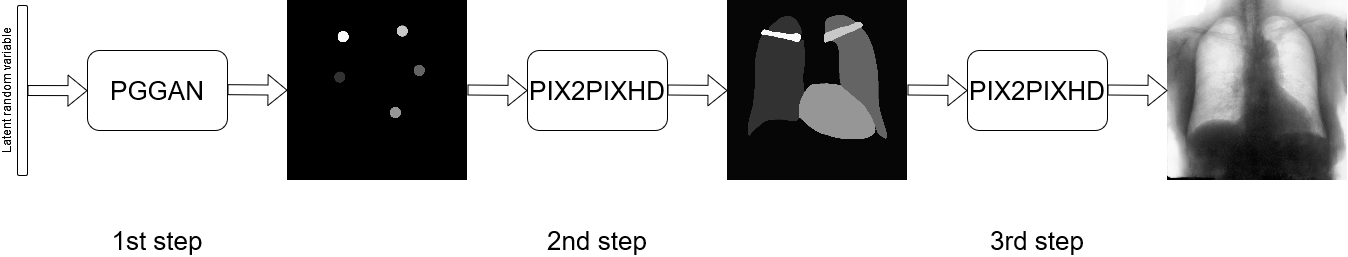}}
\caption{The three--stage image generation scheme. In the first step, dots are generated from a latent vector. Then, pix2pxHD translates dots into a label--map, and finally the label--map is translated into an X--ray image.}
\label{three_step_scheme}
\end{center}
\end{figure*}

It consists in further subdividing the generation procedure, with a first phase consisting in generating the position and type of the objects that will be generated later, regardless of their shape or appearance. This is obtained by generating label--maps that contain ``dots'' in correspondence with different anatomical parts (lungs, heart, clavicles). The dots can be considered as ``seeds'', from which, through the subsequent steps, the complete label--maps are realized (second phase). Finally, in the last step, chest X--ray images are generated from the label--maps.
The exact procedure is described in the following.
Initially, label--maps containing ``dots'', with a specific value for each anatomic part, are created. The position of the ``dot'' center is given by the centroid of each labeled anatomic part. The label--maps generated in this phase have a low resolution ($64\times{64}$), as a high level of detail is not necessary, being the exact object shapes not defined --- but only their centroid positions. It should be observed that this also allows to significantly reduce the computational burden of this stage and speedup the computation. The generated label--maps must be subsequently resized to the original image resolution --- required in the following stages of generation (a nearest neighbour interpolation has been used to maintain the original label codes) --- and translated into labels, which will be finally translated into images, using Pix2PixHD (see Figure \ref{three_step_scheme}).

\subsection{Segmentation Multiscale Attention Network} \label{SMANET}

\begin{figure*}[!ht]

\vskip 0.2in
\begin{center}
\centerline{\includegraphics[scale=0.51]{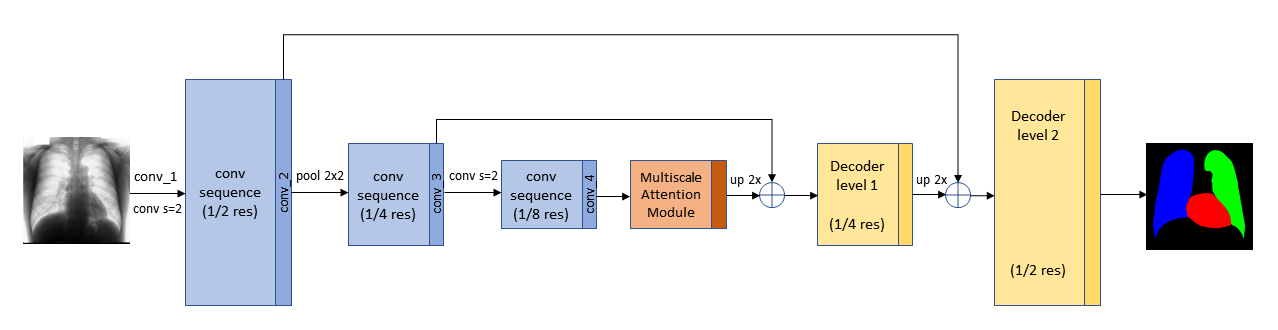}}
\caption{Scheme of the SMANet segmentation network.}
\label{SMANet_arch}
\end{center}
\end{figure*}

In this paper, the Segmentation Multiscale Attention Network (SMANet) \cite{bonechi2020weak} has been employed. The SMANet is composed by three main components, a ResNet encoder, a multi--scale attention module, and a convolutional decoder (see Figure \ref{SMANet_arch}). 
This architecture, initially proposed for scene text segmentation, is based on the Pyramid Scene Parsing Network (PSPNet) \cite{zhao2017pyramid}, a deep fully convolutional neural network with a ResNet \cite{he2016deep} encoder. Dilated convolutions (i.e. atrous convolutions \cite{papandreou2014untangling}) are used in the Resnet backbone, to widen the receptive field of the neural network in order to avoid an excessive reduction of the spatial resolution due to down--sampling. The most characteristic part of the PSPNet architecture is the pyramid pooling module (PSP), which is employed to capture features at different scale in the image. In the SMANet, the PSP module is replaced with a multi--scale attention mechanism to better focus on the relevant objects present in the image. Finally, a two--level convolutional decoder is added to the architecture to improve the recognition of small objects.

\subsection{Training Details} \label{training}
The PGGAN architecture, proposed in \cite{karras2017progressive}, has been employed for image generation; the number of parameters have been modified to speed up learning and reduce overfitting. More specifically, the maximum number of feature maps for each layer has been reduced to 64. Furthermore, since the PGGAN was used to generate seeds and labels, obtaining only the semantic label--maps in both cases, the output image has only one channel instead of three. The generation procedure (PGGAN and Pix2PixHD) has been stopped by visually examining the generated samples during the training phase. The images, generated in the various steps for all the methods, have a resolution of $1024\times{1024}$, except in the case of the ``dot'' label--maps, which, as mentioned before, are generated at a $64\times{64}$ resolution.

The SMANet is implemented in TensorFlow. Random crops of $377\times{377}$ pixels have been employed during training, whereas a sliding window of the same size has been used for testing. The Adam optimizer \cite{kingma2014adam}, based on an initial learning rate of $10^{-4}$ and a mini--batch of 17 examples, has been used to train the SMANet. 
All the experiments were carried out in a Linux environment on a single NVIDIA Tesla V100 SXM2 with 32 GB RAM.

\section{Experiments and results}   \label{experiments}
In this section, after having described the dataset on which our new proposed method was tested, we evaluate the results obtained, both qualitatively --- based on the judgment of three physicians --- and quantitatively, comparing them with related approaches present in the literature.
\subsection{Dataset}
Chest radiographs are provided by the Japanese Society of Radiological Technology (JSRT) database \cite{shiraishi2000development}. The JSRT database comprises 247 CXRs and includes images with and without lung nodules. 
All images have a resolution of $2048\times{2048}$ pixels and a spatial resolution of .175 mm/pixel, with 12 bit gray levels.
Instead, segmentation supervisions for the JSRT database are available in the Segmentation in Chest Radiographs (SCR) dataset \cite{van2006segmentation}. More precisely, this dataset provides chest X--ray supervisions which correspond with the pixel--level positions of the different anatomical parts. Such  supervisions were produced by two observers who segmented five objects in each image: the two lungs, the heart and the two clavicles. The first observer was a medical student and his segmentation was used as the gold standard, while the second observer was a computer science student, specialized in medical imaging, and his segmentation was considered that of a human expert.

The SCR dataset comes with an official splitting, which is employed in this paper and consists of 124 images for learning and 123 for testing. We  use two different experimental configurations. 
In the former, called FULL\_DATASET, all the training images are exploited. More precisely, the PGGAN generation network is trained on the basis of 744 images, available in the SCR training set and obtained with the augmentation procedure described above. The SMANET is trained on 7500 synthetic images, generated by the PGGAN, and fine--tuned on the 744 images extracted from the SCR training set, while 2500 synthetic images are used for validation. For the second configuration, called TINY\_DATASET,
only a 10\% of the SCR training set is used and the PGGAN is trained on only 66 images (obtained both from SCR and with augmentation); instead, the SMANET is trained exactly as above, except for the fine--tuning, which is carried out on 66 images.

\subsection{Quantitative results}
Generated images have been employed to train a deep semantic segmentation network. The rationale behind the approach is that the performance of the network trained on the generated data reflects the data quality and variety. 
A good performance of the segmentation network indicates that the generated data successfully capture the true distribution of the real samples. 
To assess the segmentation results, some standard evaluation metrics have been used. The Jaccard Index, $J$, also called Intersection Over Union (IOU),  measures the similarity between two finite sample sets --- the predicted segmentation and the target mask in this case ---, and is defined as the size of their intersection divided by the size of their union. 
For binary classification, the Jaccard index can be framed in the following formula:
$$
    J = \frac{TP}{TP + FP + FN}
$$
where $TP,FP,FN$ denote the number of true positives, false positives 
and false negatives, respectively.
Instead, the Dice Score, $DSC$, is defined as:
$$
    DSC = \frac{2\times TP}{2\times TP + FP + FN}
$$
$DSC$ is a \textit{quotient of similarity between sets} and ranges between 0 and 1. 

The experiments can be divided into two phases: first, we evaluate the generation procedure described in Section \ref{generation:three_step} using the FULL\_DATASET, then, we compare this approach with the other two methods described in Sections \ref{generation:single_step} and \ref{generation:two_step} using the TINY\_DATASET.
The purpose of this latter experiment is to evaluate whether multi--stage generation methods are actually more effective in producing data suitable for semantic segmentation with a limited amount of data. In particular, in the experimental setup based on the FULL\_DATASET, for the three--stage method, the generation network has been trained on all the SCR training images, to which the augmentation procedure described in Section \ref{generation} has been applied. Then, 10,000 synthetic images have been generated and used to train the semantic segmentation network. 
Moreover, we evaluated  a fine--tuning of the network on the SCR real images 
after the pre--training on the generated images.
The results, shown in Table \ref{tab:7500_training}, are compared with those obtained using only real images to train the semantic segmentation network, which can be considered as a baseline. 

Next, the TINY\_DATASET has been used in order to evaluate the performance of the methods with a very small dataset. More precisely, the following experimental setups, whose results are shown in Table \ref{tab2},  are considered:
\begin{itemize}
    \item REAL -- only real images are used for training the semantic segmentation network;
    \item SINGLE--STAGE -- the segmentation network uses the images generated by the single--stage method (Synth 1 in the tables) for training while real images are employed for fine--tuning (Finetune in the tables);
    \item TWO--STAGES -- the images generated with the two--stage method are used to pre--train the segmentation network (Synth 2) while real images are used for fine--tuning;
    \item THREE--STAGES -- the images generated with the three--stage method are used for training the segmentation network (Synth 3), while real images are employed for fine--tuning.
\end{itemize}   
In this case, the PGGAN has been trained on 66 images, based on 11 images randomly chosen from the entire training set to which the augmentation described above has been applied.

In general, we can see that the best results are obtained with the three--stage method followed by fine--tuning.
From Table \ref{tab:7500_training}, we observe a small improvement in results using a fine--tune on a network previously trained with images generated using the three--stage method. Therefore, the three--stage method provides good synthetic data, but the advantage given by generated images is low when the training set is large.
 Conversely, when  few training images are available, in the TINY\_DATASET setup, multi--stage methods outperform the baseline (column REAL of Table \ref{tab2}) and this happens even without fine--tuning. Thus, in this case, the advantage provided by synthetic images is evident. Moreover, the three--stage method outperforms the two--stage approach, even with fine--tuning, which confirms our claim that splitting the generation procedure may provide a performance increase when few training images are available.

Finally, it is worth noting that fine--tuning improves the performance  
of the three--stage method,  both in the FULL\_DATASET and in the TINY\_DATASET framework, which does not hold for the two--stage method. This behaviour 
may be explained by some complementary information that is captured from real images only with the three--stage method. Actually, we may argue that, in different phases
of a multi--stage approach, different types of information can be captured: 
such a diversification seems to provide an advantage to the three--stage method, 
which develops some capability to model the data domain with more orthogonal information. 

\begin{table}[t]
\begin{center}
\caption{Evaluation of the proposed methods based on the FULL\_DATASET, using 2500 generated images for the validation set. {\bf REAL} corresponds to the results obtained using the official training set; \textit{Synth 3} corresponds to the results obtained using only the generated images, while in the \textit{FINETUNE}
column, real data are employed for fine-tuning. \label{tab:7500_training}}
\vspace{0.3cm}
\begin{tabular}{ccccc}
     \toprule
     & & \multirow{2}{*}{\textbf{REAL}} & \multicolumn{2}{c}{\textbf{THREE--STAGE}} \\
     \cmidrule(lr){4-5}
     & & &  \textit{Synth 3} & \textit{FINETUNE} \\
     \midrule
     \multirow{4}{*}{\textbf{J}} & Left Lung & 96.10 & 95.30 & \textbf{96.22} \\
& Heart & 90.78 & 87.25 & \textbf{91.11} \\
& Right Lung & \textbf{96.85} & 96.15 & 96.79 \\
& \textbf{Average} & 94.58 & 92.90 & \textbf{94.71} \\
\midrule
\multirow{4}{*}{\textbf{DSC}} & Left Lung & 98.01 & 97.6 & \textbf{98.07} \\
& Heart & 95.17 & 93.19 & \textbf{95.35} \\
& Right Lung & \textbf{98.40} & 98.04 & 98.37 \\
& \textbf{Average} & 97.19 & 96.28 & \textbf{97.26} \\
\bottomrule

\end{tabular}
\end{center}
\end{table}

\begin{table}[t]
\begin{center}
\caption{Evaluation of the proposed methods based on the TINY\_DATASET, using 2500 generated images for the validation set. {\bf REAL} corresponds to the results obtained using the official training set;  \textit{Synth 1}, \textit{Synth 2}, \textit{Synth 3}, correspond to the results obtained using only the generated images, while in the \textit{FINETUNE} columns, real data are employed for fine–tuning. \label{10_7500_training}
}
\vspace{0.3cm}
\begin{tabular}{ccccccccc}
    \toprule 
     &  & \multirow{2}{*}{\textbf{REAL}} & \multicolumn{2}{c}{\textbf{SINGLE--STAGE}} & \multicolumn{2}{c}{\textbf{TWO--STAGE}} & \multicolumn{2}{c}{\textbf{THREE--STAGE}} \\
     \cmidrule(lr){4-5} \cmidrule(lr){6-7} \cmidrule(lr){8-9} 
  & & & \textit{Synth 1} & \textit{FINETUNE} & \textit{Synth 2} & \textit{FINETUNE} & \textit{Synth 3} & \textit{FINETUNE} \\
  \midrule
  \multirow{4}{*}{\textbf{J}} & Left Lung & 93.70 & 55.59 & 74.11 & 94.91 & 94.4 & 94.96 & \textbf{95.29} \\
& Heart & 85.50 & 0.07 & 37.47 & 86.98 & 85.21 & 87.27 & \textbf{87.47} \\
& Right Lung & 93.70 & 52.78 & 79.99 & 95.90 & 95.44 & 95.90 & \textbf{95.92} \\
& \textbf{Average} & 90.97 & 36.15 & 63.86 & 92.60 & 91.68 & 92.71 & \textbf{92.89} \\
\midrule
\multirow{4}{*}{\textbf{DSC}} & Left Lung & 96.75 & 71.46 & 85.13 & 97.39 & 97.12 & 97.42 & \textbf{97.59} \\
& Heart & 92.18 & 0.13 & 54.51 & 93.04 & 92.02 & 93.20 & \textbf{93.32} \\ 
& Right Lung & 96.74 & 69.09 & 88.89 & 97.91 & 97.66 & 97.90 & \textbf{97.92} \\
& \textbf{Average} & 95.22 & 46.89 & 76.18 & 96.11 & 95.60 & 96.17 & \textbf{96.28} \\
  \bottomrule
\end{tabular}
\label{tab2}
\end{center}
\end{table}

\subsection{Comparison with other approaches}
Table~\ref{comparison} shows our best results and the segmentation performance published by all recent methods, of which we are aware, on the SCR dataset. According to the results in the table, the three--stage method obtained the best performance score both for the lungs and the heart.

However, it is worth mentioning that Table \ref{comparison} gives only a rough idea of the state--of--the--art, since  a direct comparison between the proposed method and other approaches is not feasible,
being our primary focus on image generation, in contrast with the comparative approaches that are mainly devoted to segmentation,
and for which no results are reported on small image datasets. Moreover, the previous methods used different partition of the SCR dataset to obtain the training and the test set, such as 2--fold, 3--fold, 5--fold cross--validation or ad hoc splittings, which are often not publicly available,
while, in our experiments, we preferred to use the original partition, provided with the SCR dataset\footnote{Note that, compared to most of the
other solutions used in comparative methods, the original subdivision has the disadvantage of producing a smaller training set, 
which is not in conflict, however, with the purpose of the present work.}. Finally,  also a variety of different image size have been used, ranging from $256\times256$, to $400\times400$, and to  $512\times512$ --- the resolution used in this work.

\begin{table}[ht]
\begin{center}
\small
\caption{Comparison of segmentation results among different methods on the SCR dataset (CV stands for cross--validation). \label{comparison}
}
\vspace{0.3cm}
\begin{tabular}{cccccccc}
    \toprule 
     \multirow{2}{*}{\textbf{METHOD}} & \multirow{2}{*}{\textbf{IMAGE SIZE}} & \multirow{2}{*}{\textbf{AUGMENTATION}} & \multirow{2}{*}{\textbf{EVALUATION SCHEME}} & 
     \multicolumn{2}{c}{\textbf{LUNGS}} & \multicolumn{2}{c}{\textbf{HEART}} \\
     \cmidrule(lr){5-6} \cmidrule(lr){7-8}  
  & & & & \textit{DSC} & \textit{J} & \textit{DSC} & \textit{J} \\
  \midrule
\textbf{Human expert}\cite{van2006segmentation} & $2048\times2048$ & No & - & - & 94.6 & - & 87.8 \\ 
\textbf{U--Net}\cite{wang2017segmentation} & $256\times256$ & No & 5-fold CV & - & 95.9 & - & 89.9 \\ 
\textbf{InvertedNet}\cite{novikov2018fully} & $256\times256$ & No & 3-fold CV & 97.4 & 95 & 93.7 & 88.2 \\ 
\textbf{SegNet}\cite{islam2018towards} & $256\times256$ & No & 5-fold CV & 97.9 & 95.5 & 94.4 & 89.6 \\ 
\textbf{FCN}\cite{islam2018towards} & $256\times256$ & No & 5-fold CV & 97.4 & 95 & 94.2 & 89.2 \\ 
\textbf{SCAN}\cite{novikov2018fully} & $400\times400$ & No & training/test split (209/38) & 97.3 & 94.7 & 92.7 & 86.6 \\ 
\textbf{Our three--stage method} & $512\times512$ & Yes & official split & \textbf{98.2} & \textbf{96.5} & \textbf{95.36} & \textbf{91.1} \\ 
  \bottomrule
\end{tabular}
\end{center}
\end{table}

\subsection{Qualitative results}
In this section, some examples  of images and corresponding segmentations, generated with the approaches described in Section \ref{generation}, are qualitatively examined. We also report some comments from three  physicians on the generated segmentations, to provide a medical assessment of the quality of our method.

\begin{figure*}[t!]

\begin{center}

% GEN THREE--STEP 
\subfloat[]{
\includegraphics[width=2.63cm,height=2.63cm]{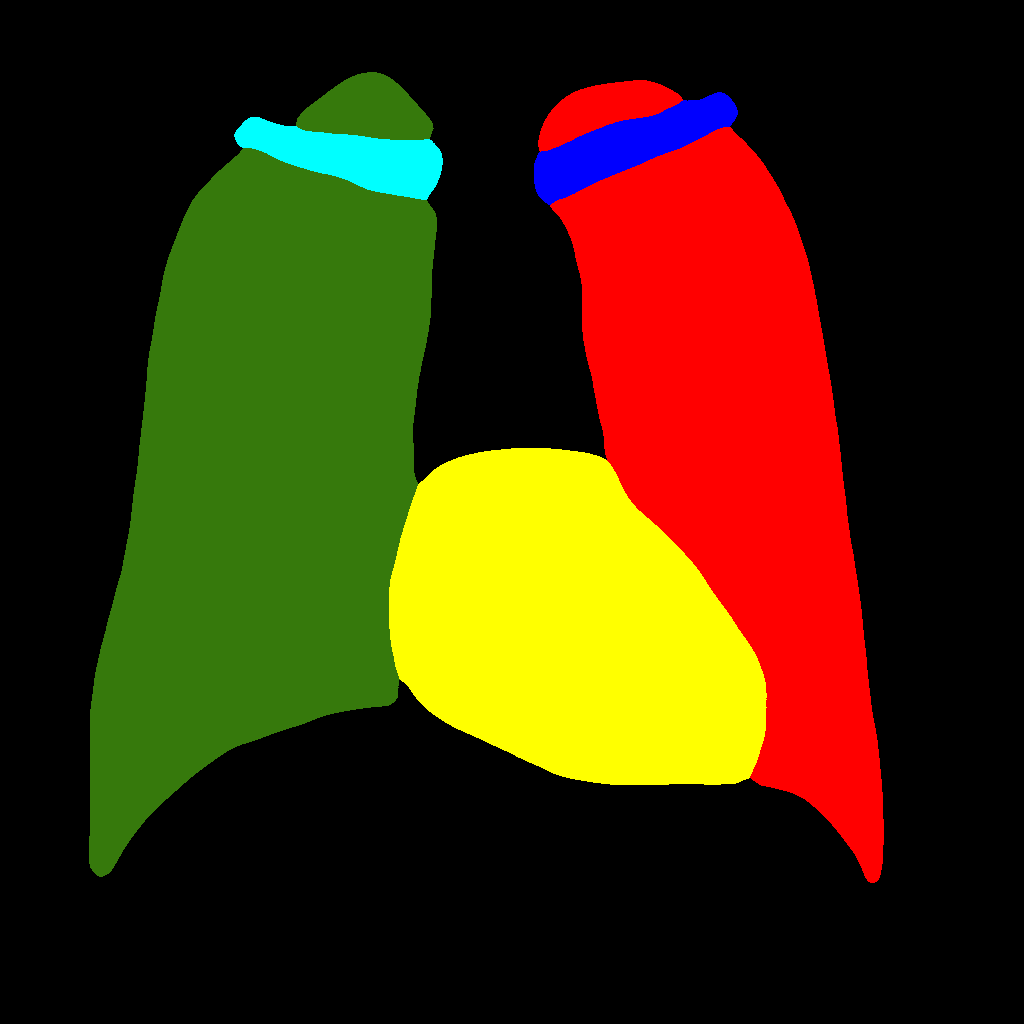}
\includegraphics[width=2.63cm,height=2.63cm]{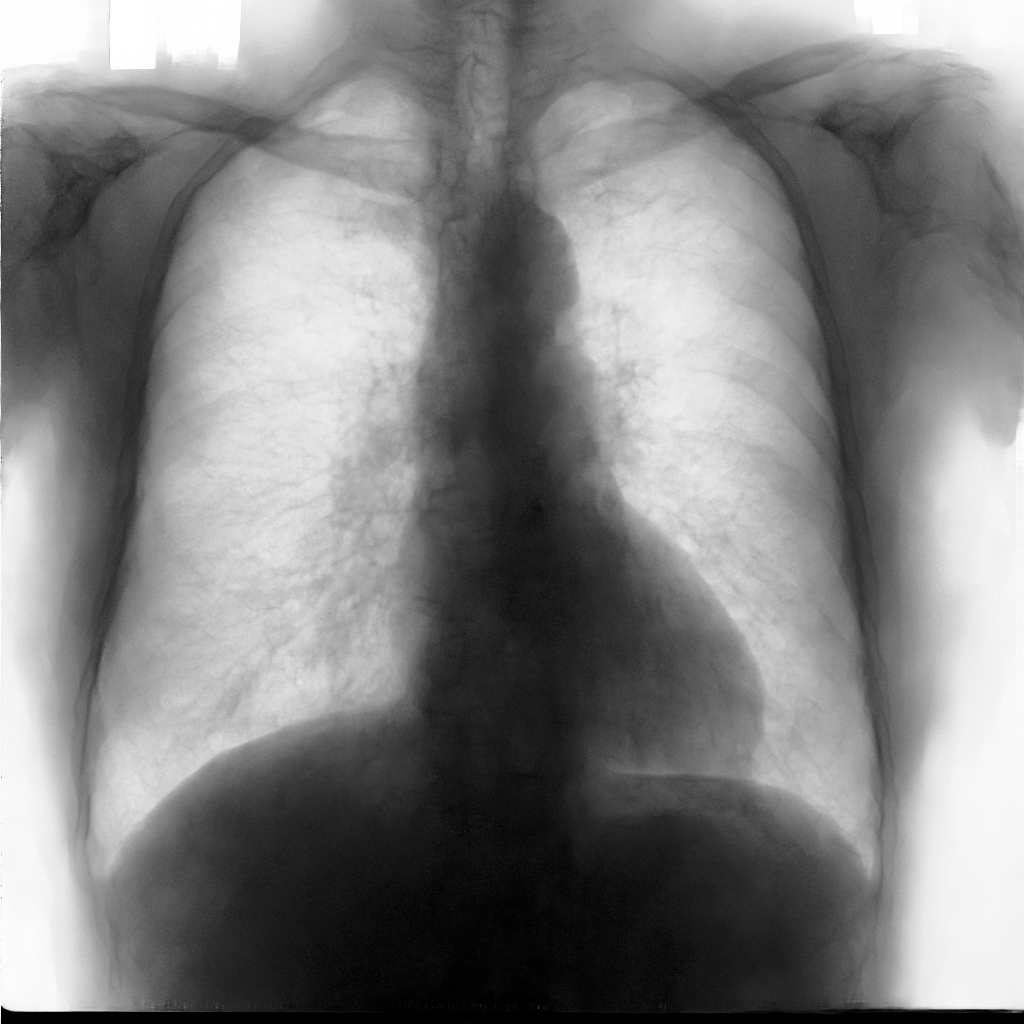} 
\includegraphics[width=2.63cm,height=2.63cm]{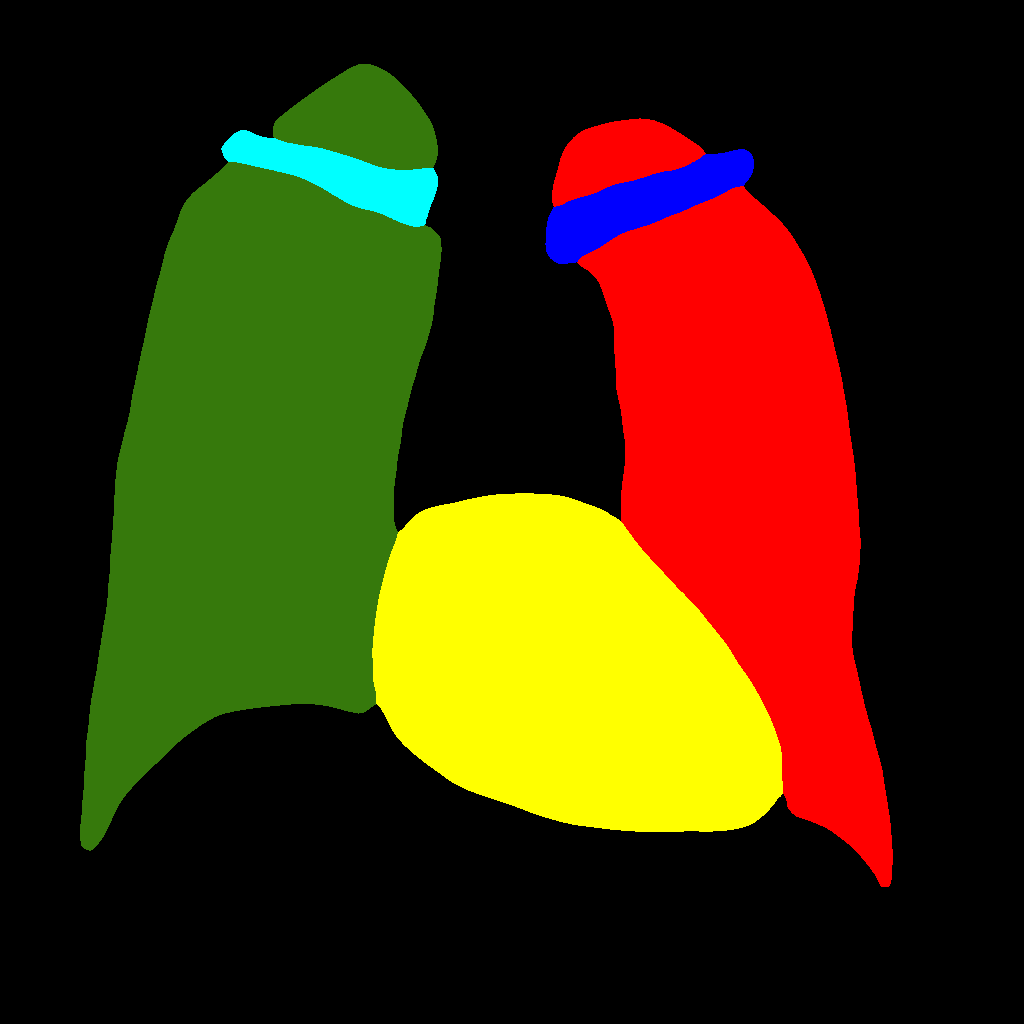}
\includegraphics[width=2.63cm,height=2.63cm]{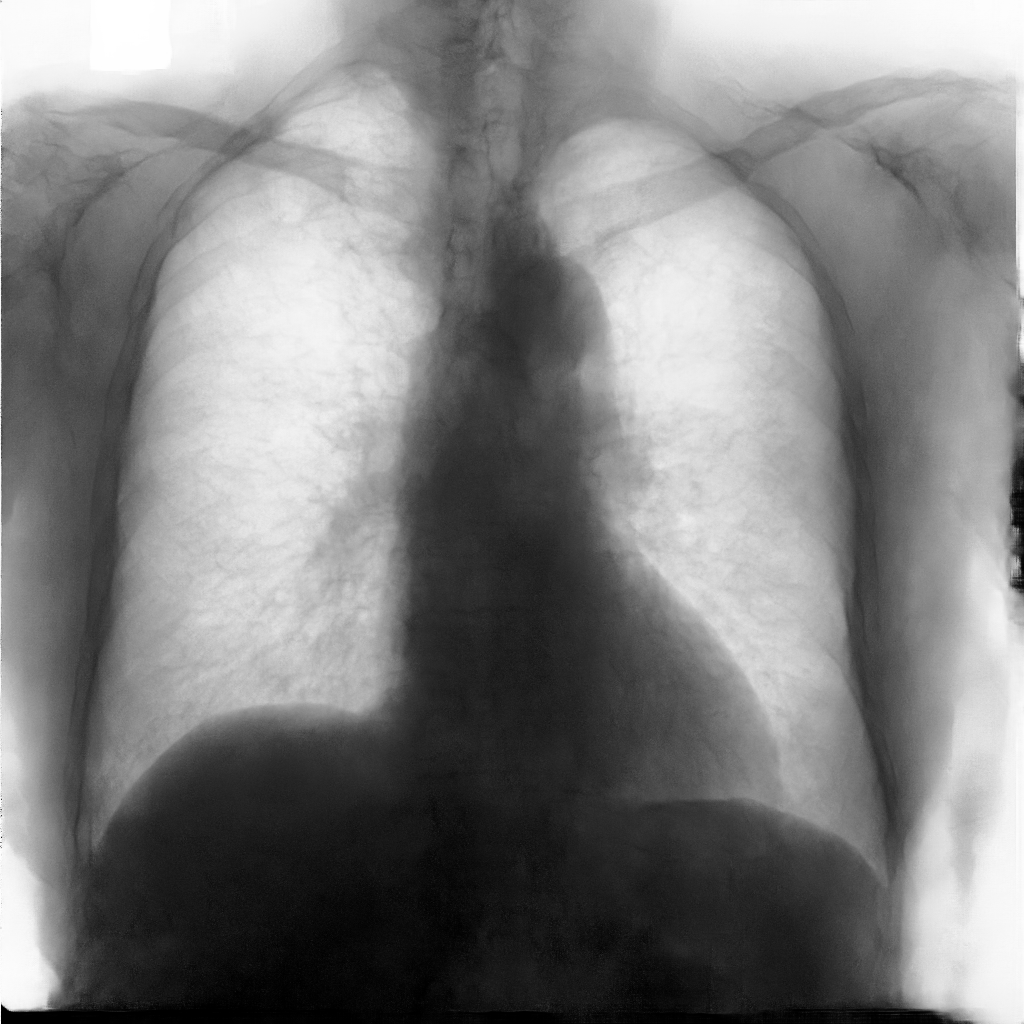}
\includegraphics[width=2.63cm,height=2.63cm]{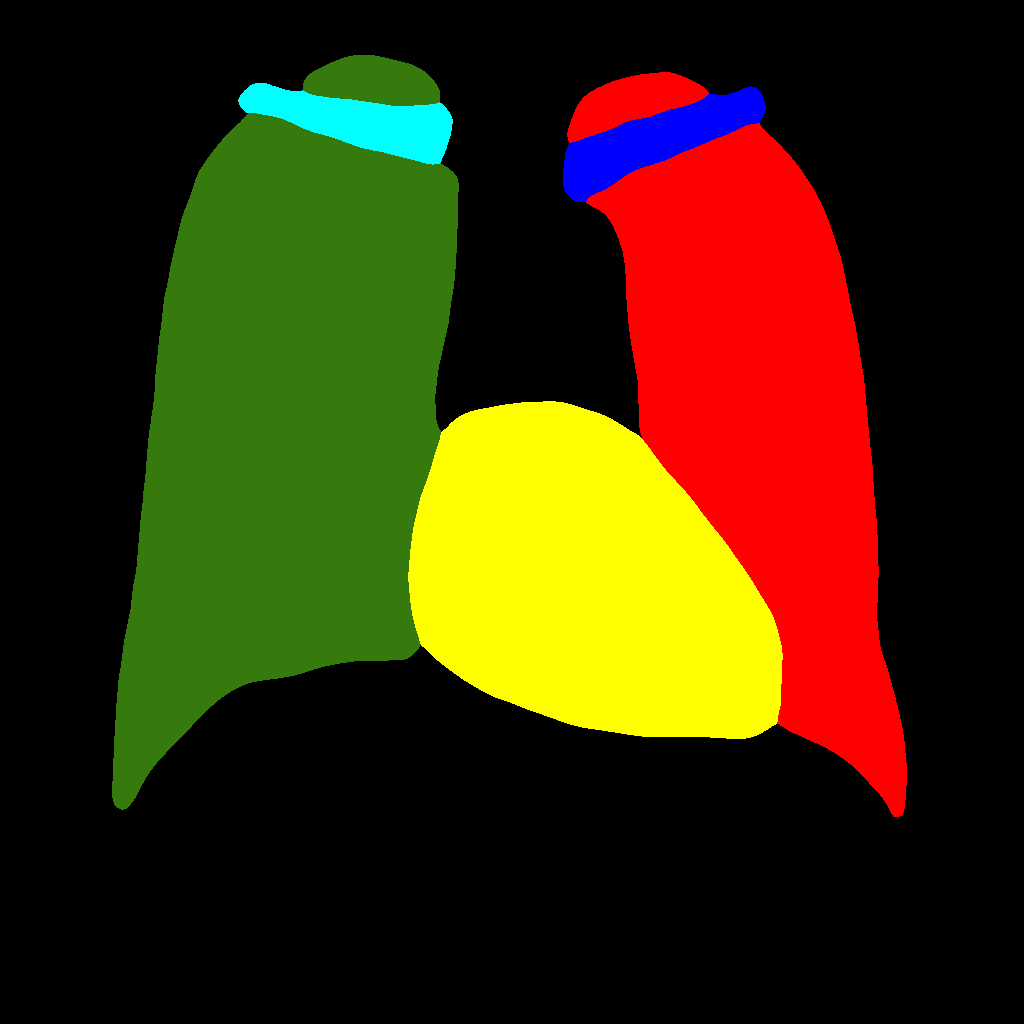}
\includegraphics[width=2.63cm,height=2.63cm]{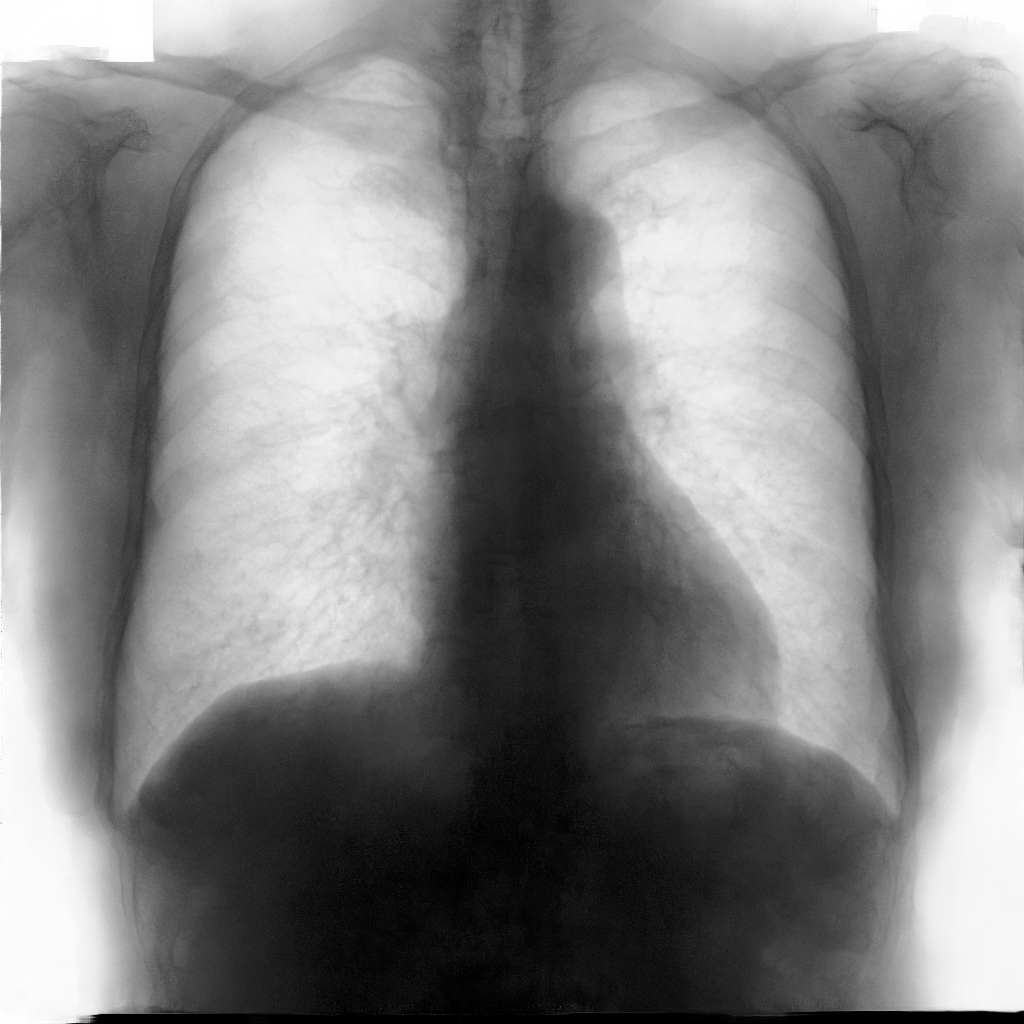}
}
\end{center}
\caption{Examples three--stage generated images based on the FULL\_DATASET.}
\label{fig:gen_examples_full}
\end{figure*}

\begin{figure*}[t!]

\vspace{0.7cm}
\begin{center}
\subfloat[Single--stage 10\% generated images.]{
% GEN SINGLE--STEP 10%
\includegraphics[width=2.63cm,height=2.63cm]{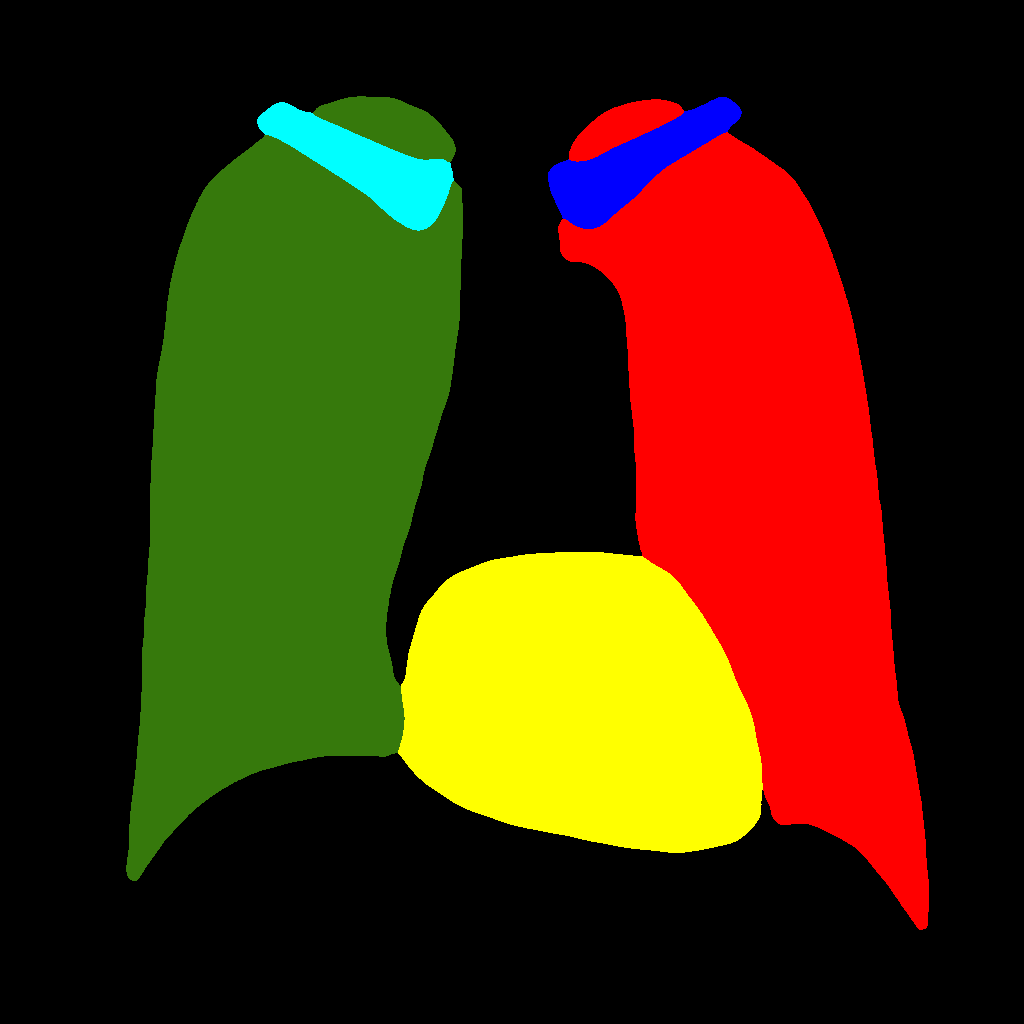}
\includegraphics[width=2.63cm,height=2.63cm]{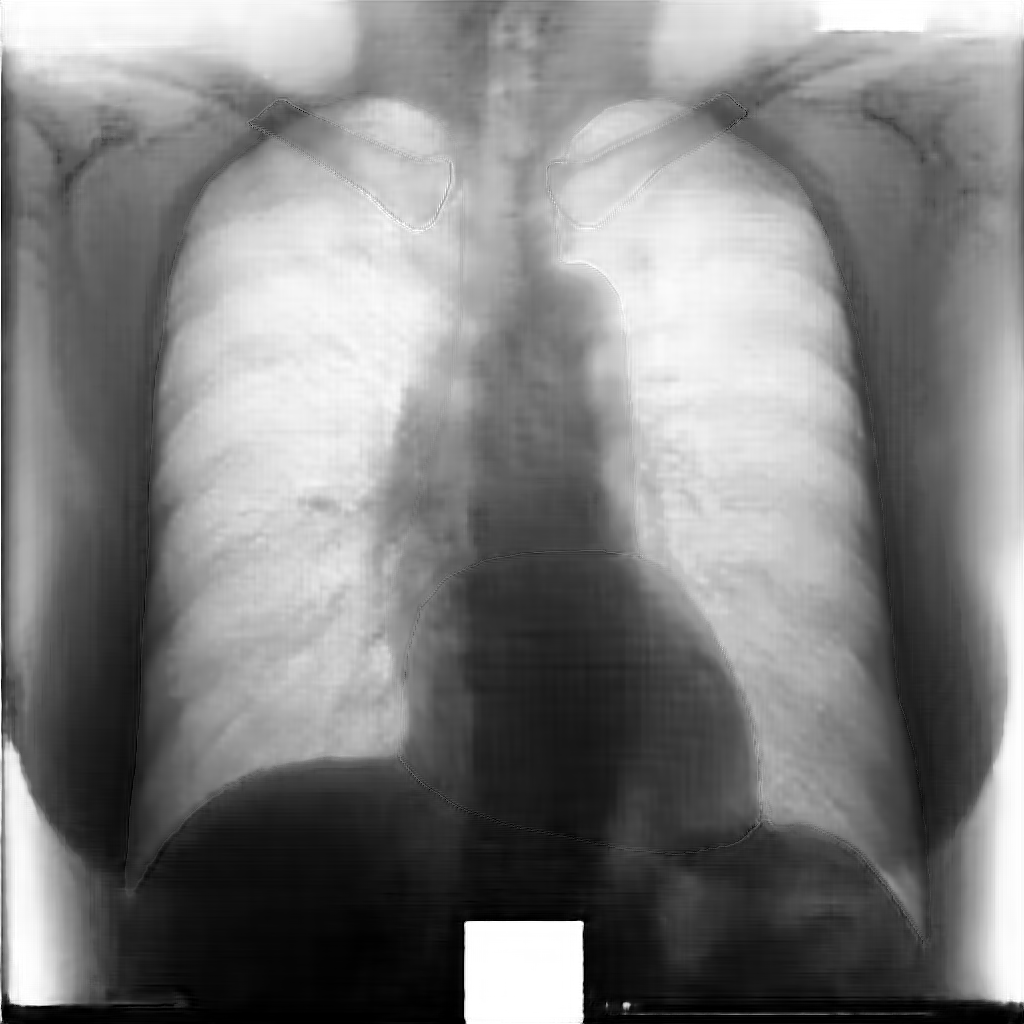} 
\includegraphics[width=2.63cm,height=2.63cm]{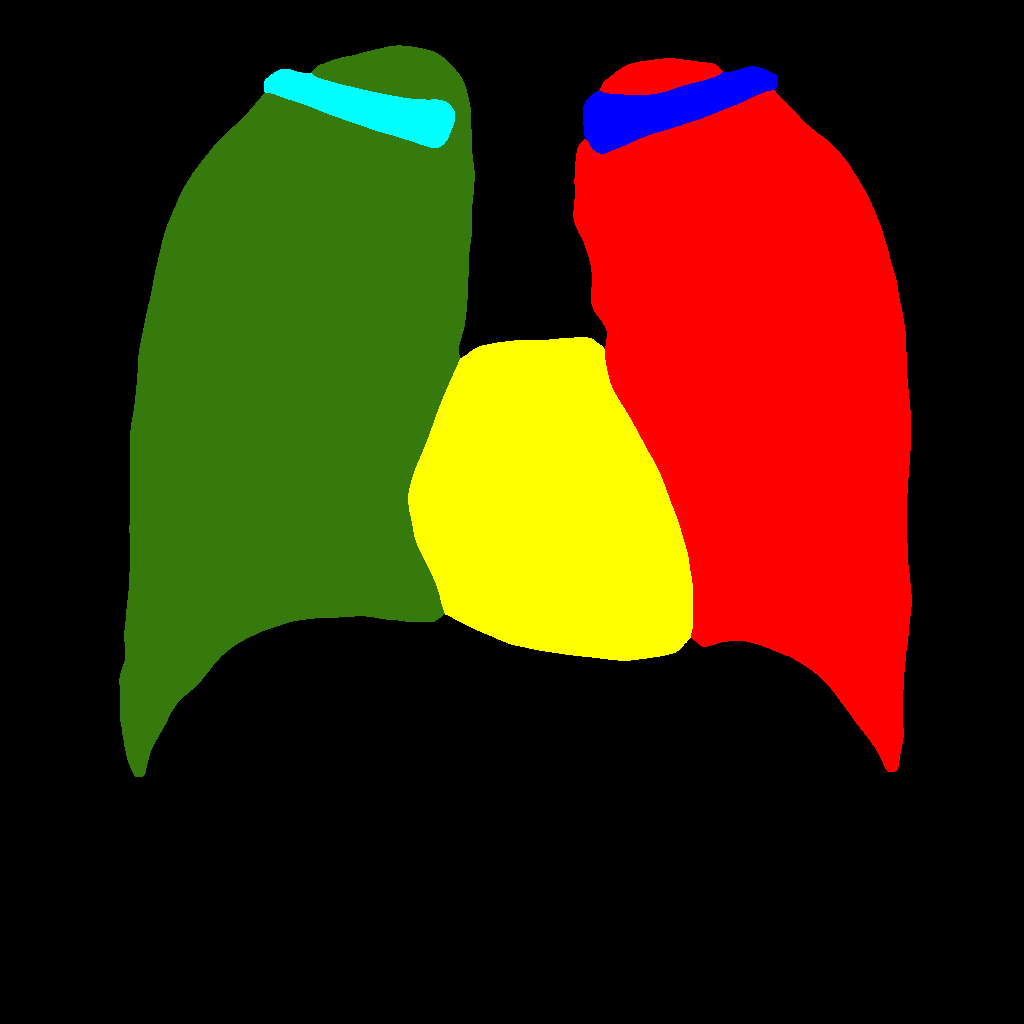}
\includegraphics[width=2.63cm,height=2.63cm]{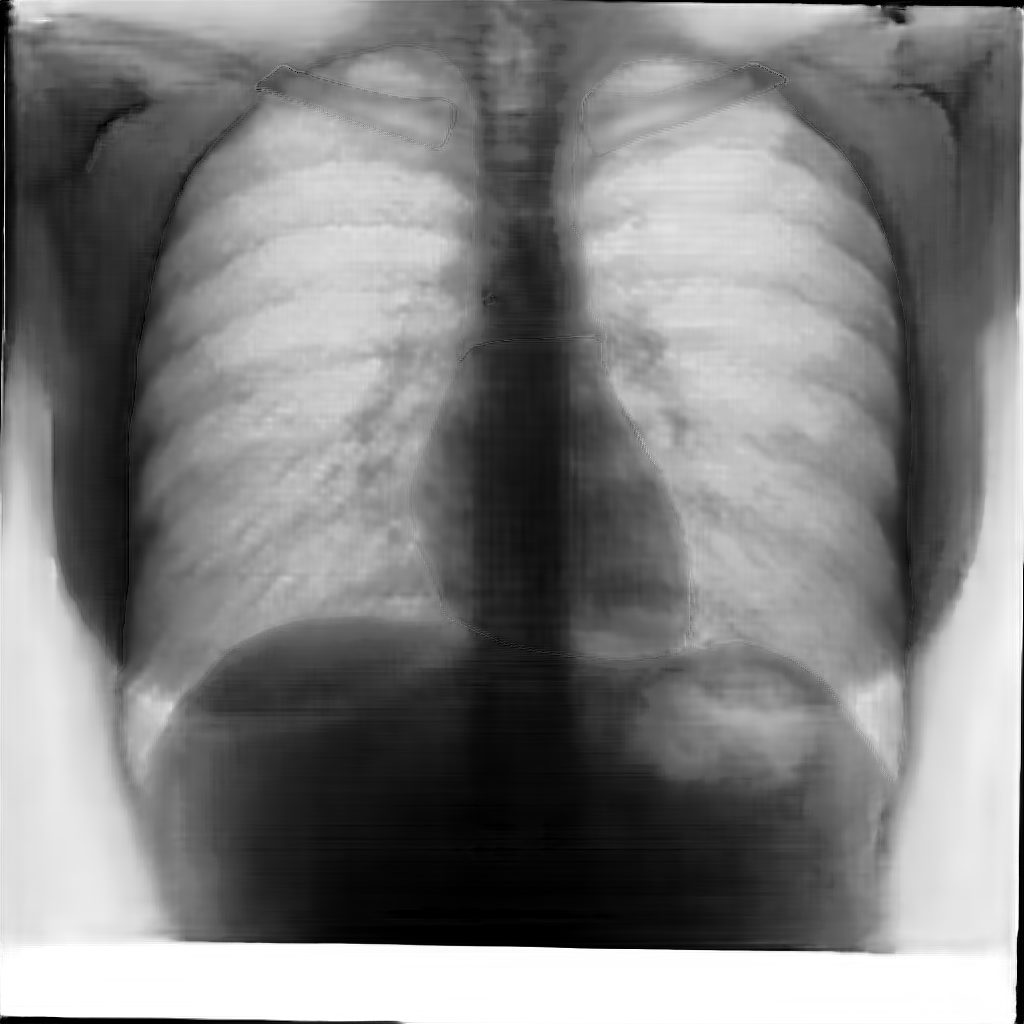}
\includegraphics[width=2.63cm,height=2.63cm]{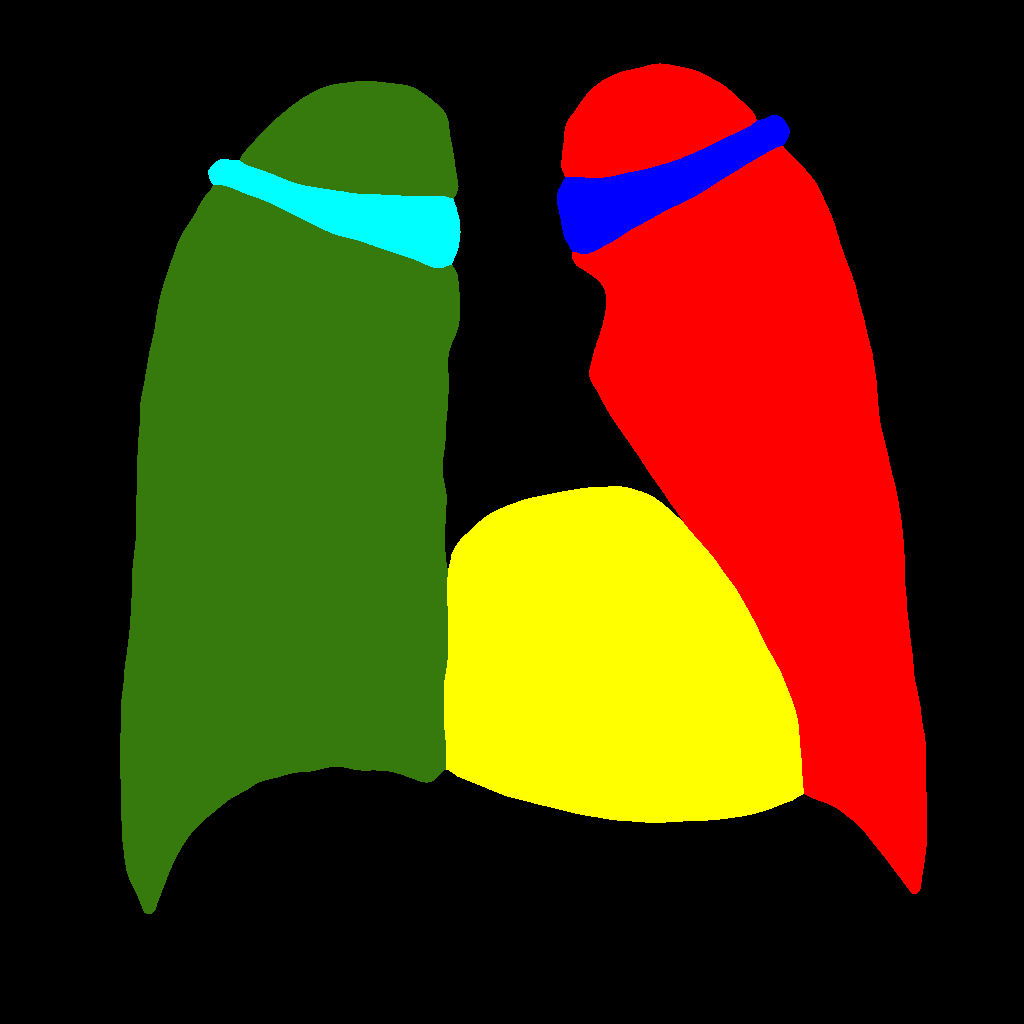}
\includegraphics[width=2.63cm,height=2.63cm]{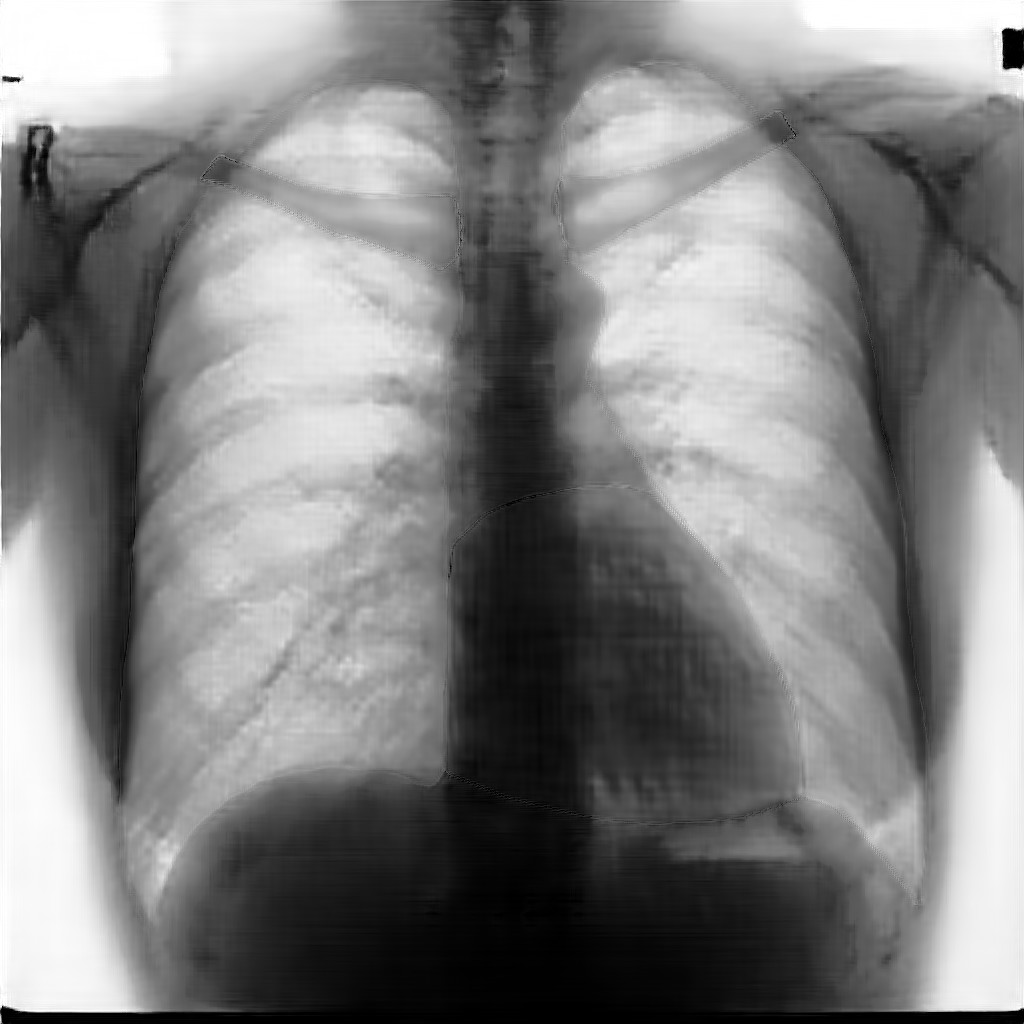}
} \\
\subfloat[Two--stage 10\% generated images.]{
% GEN TWO--STEP 10%
\includegraphics[width=2.63cm,height=2.63cm]{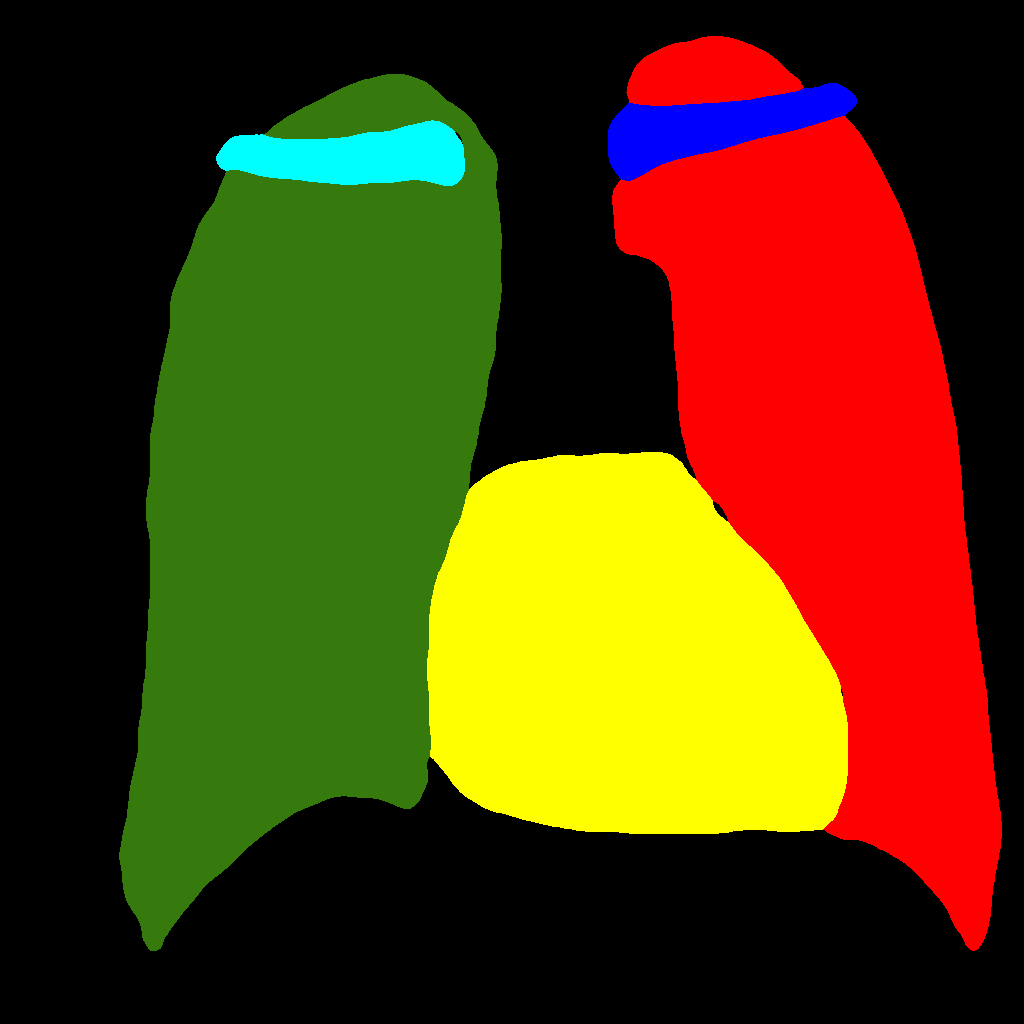}
\includegraphics[width=2.63cm,height=2.63cm]{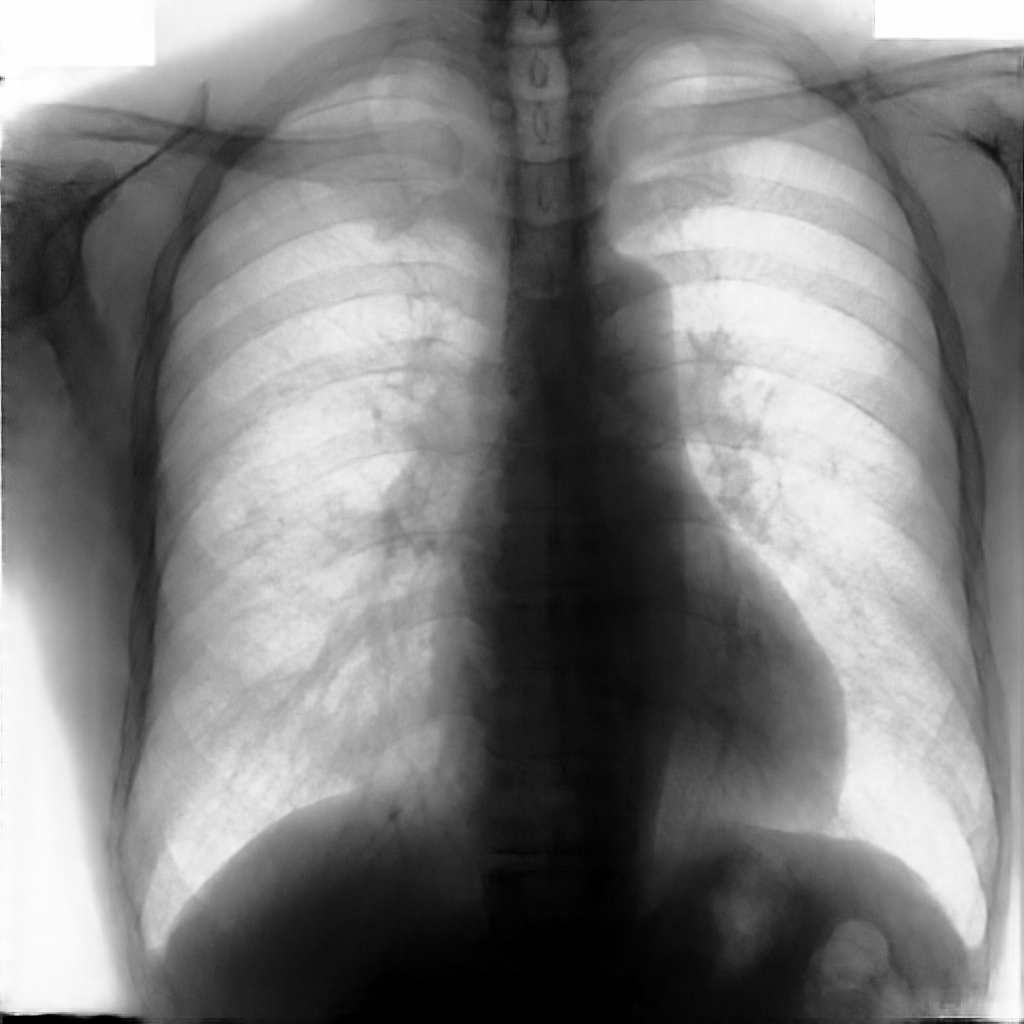} 
\includegraphics[width=2.63cm,height=2.63cm]{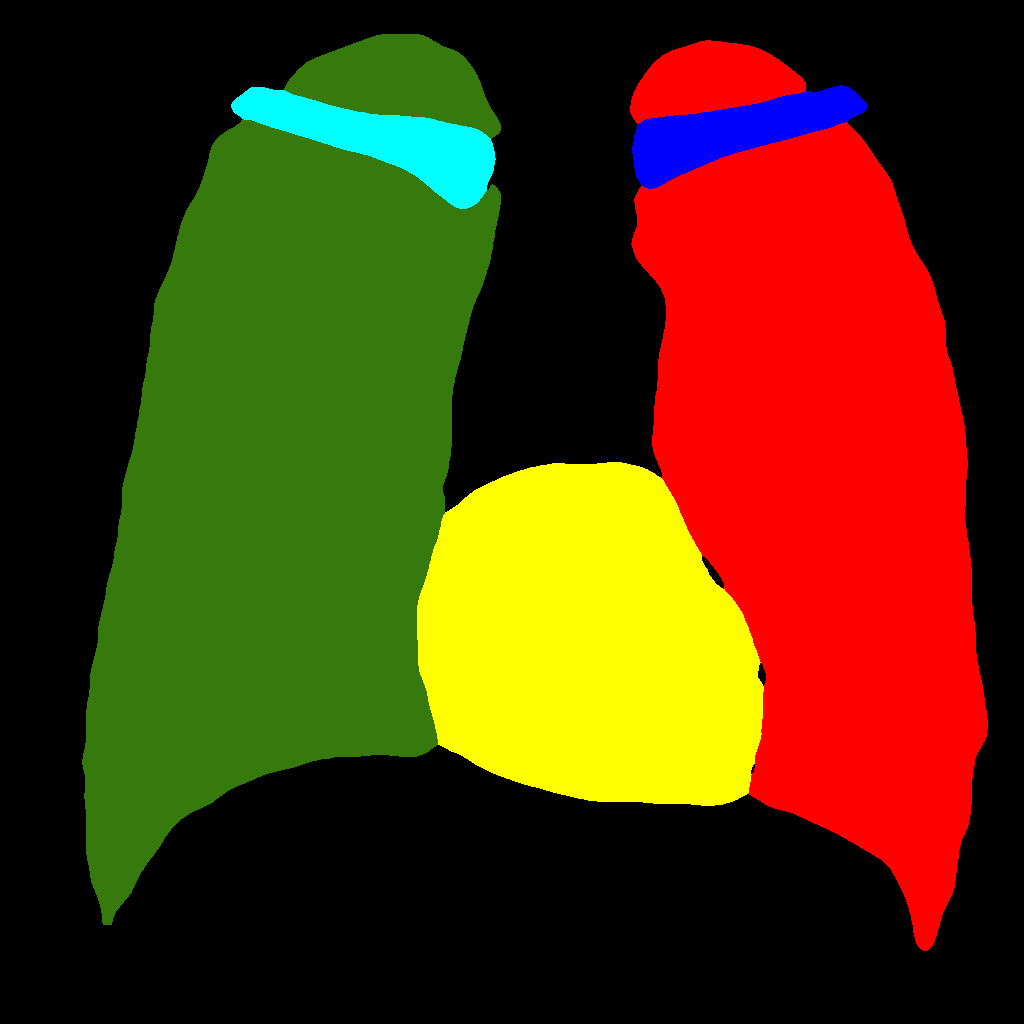}
\includegraphics[width=2.63cm,height=2.63cm]{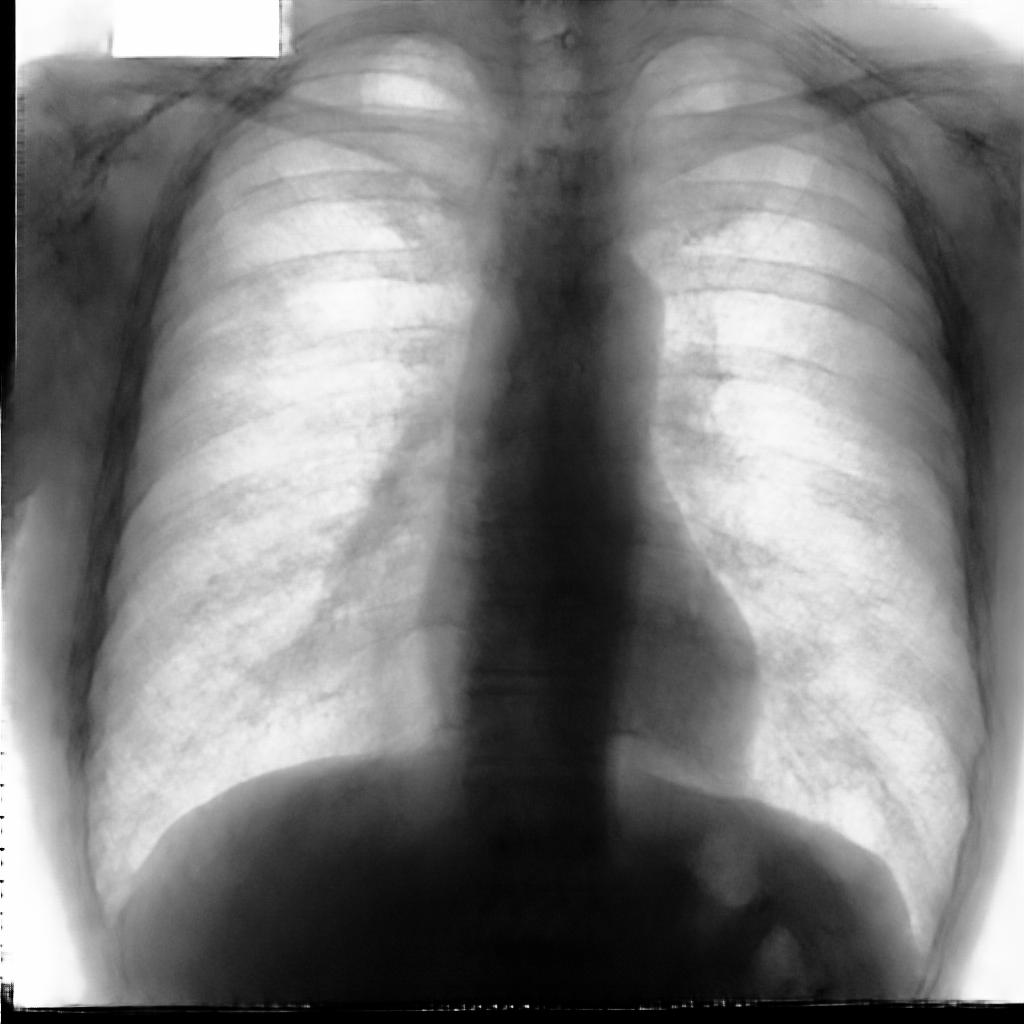}
\includegraphics[width=2.63cm,height=2.63cm]{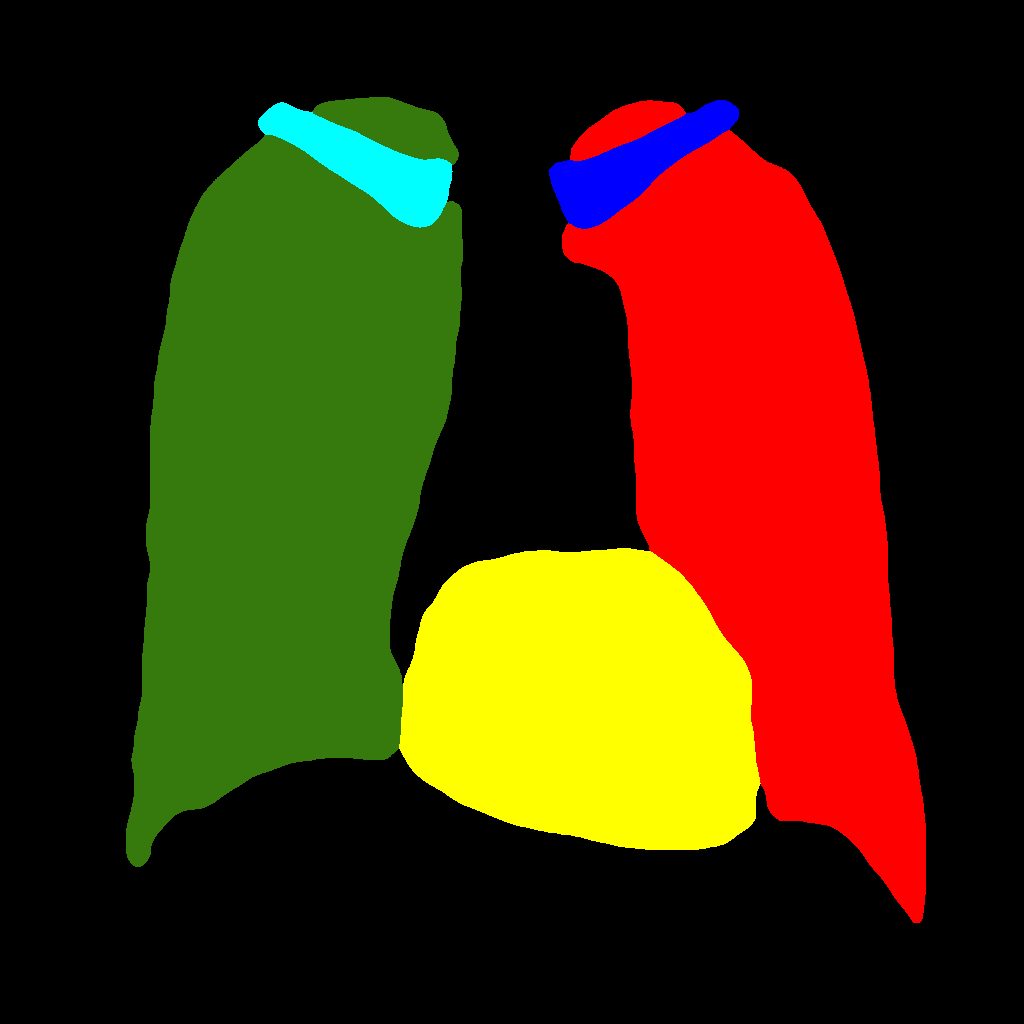}
\includegraphics[width=2.63cm,height=2.63cm]{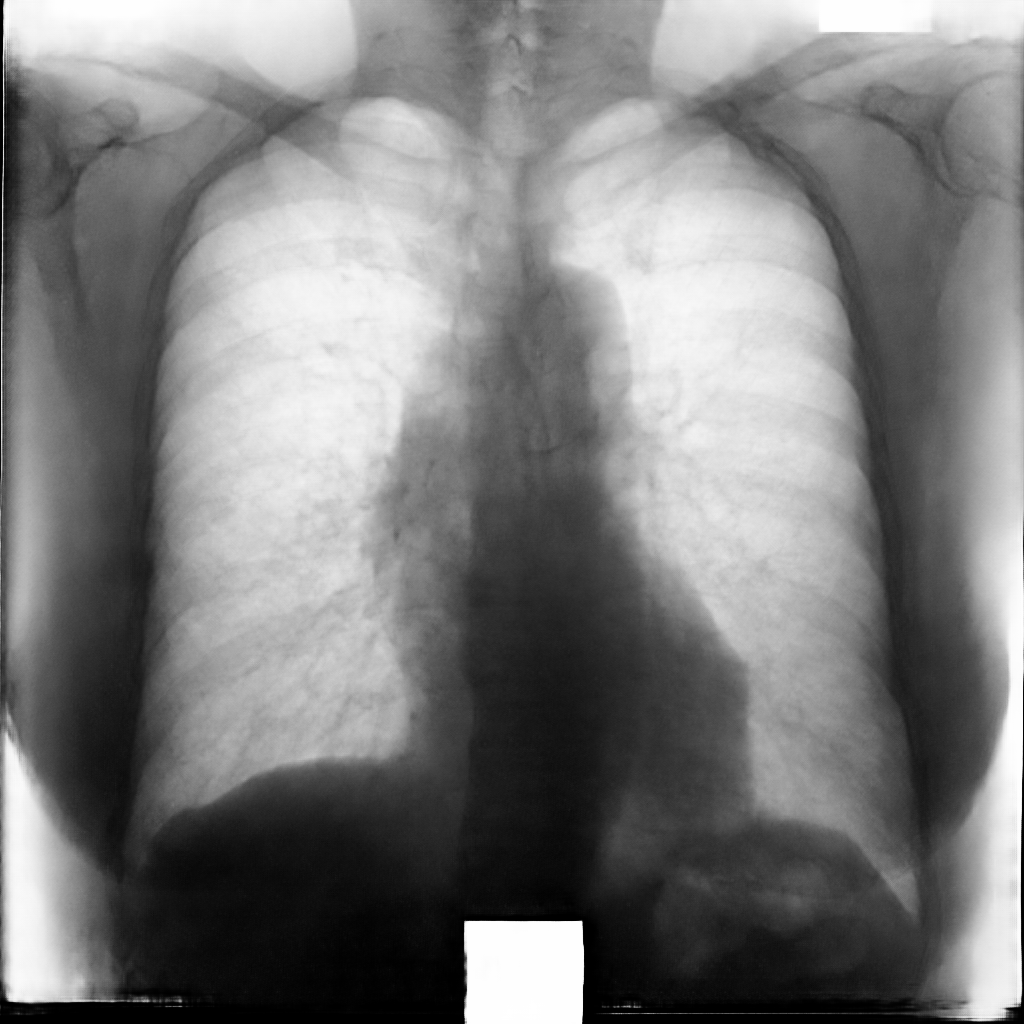}
} \\
\subfloat[Three--stage 10\% generated images.]{
% GEN THREE--STEP 10%
\includegraphics[width=2.63cm,height=2.63cm]{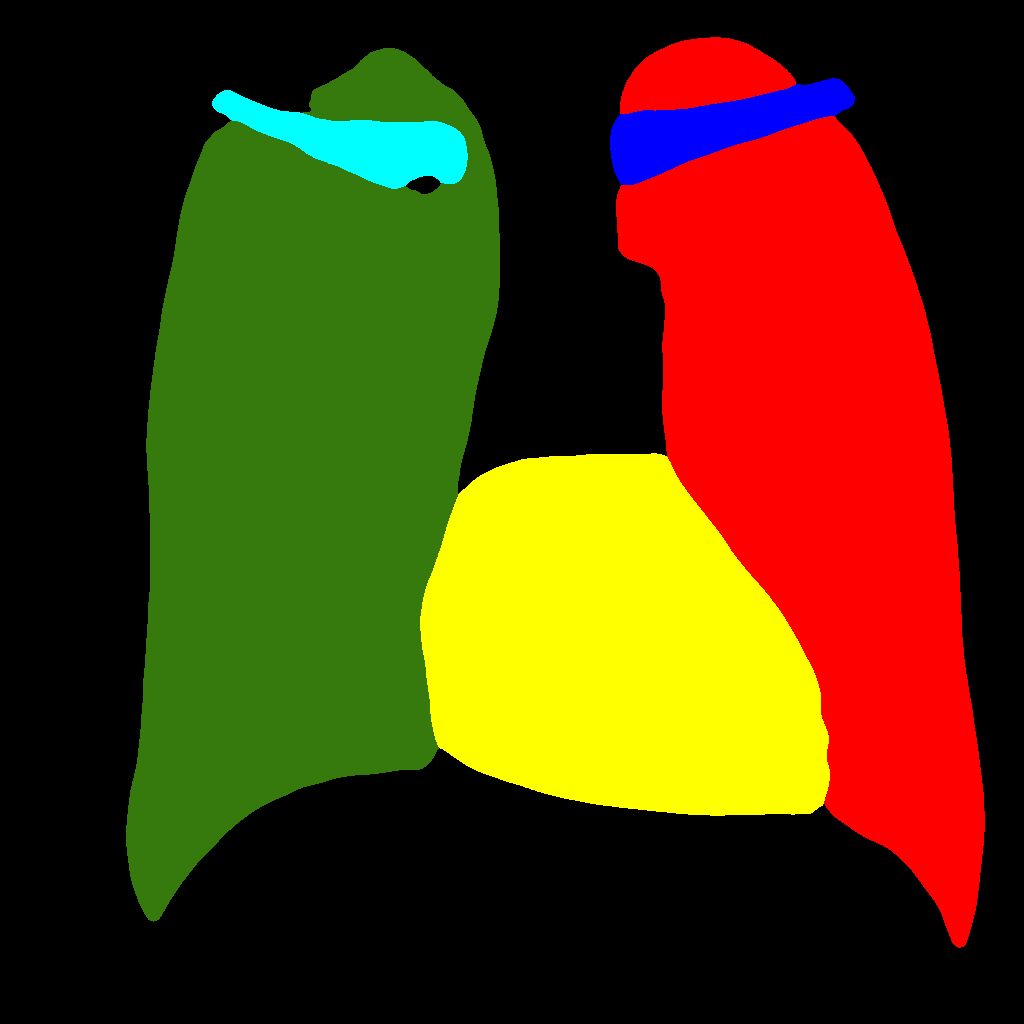}
\includegraphics[width=2.63cm,height=2.63cm]{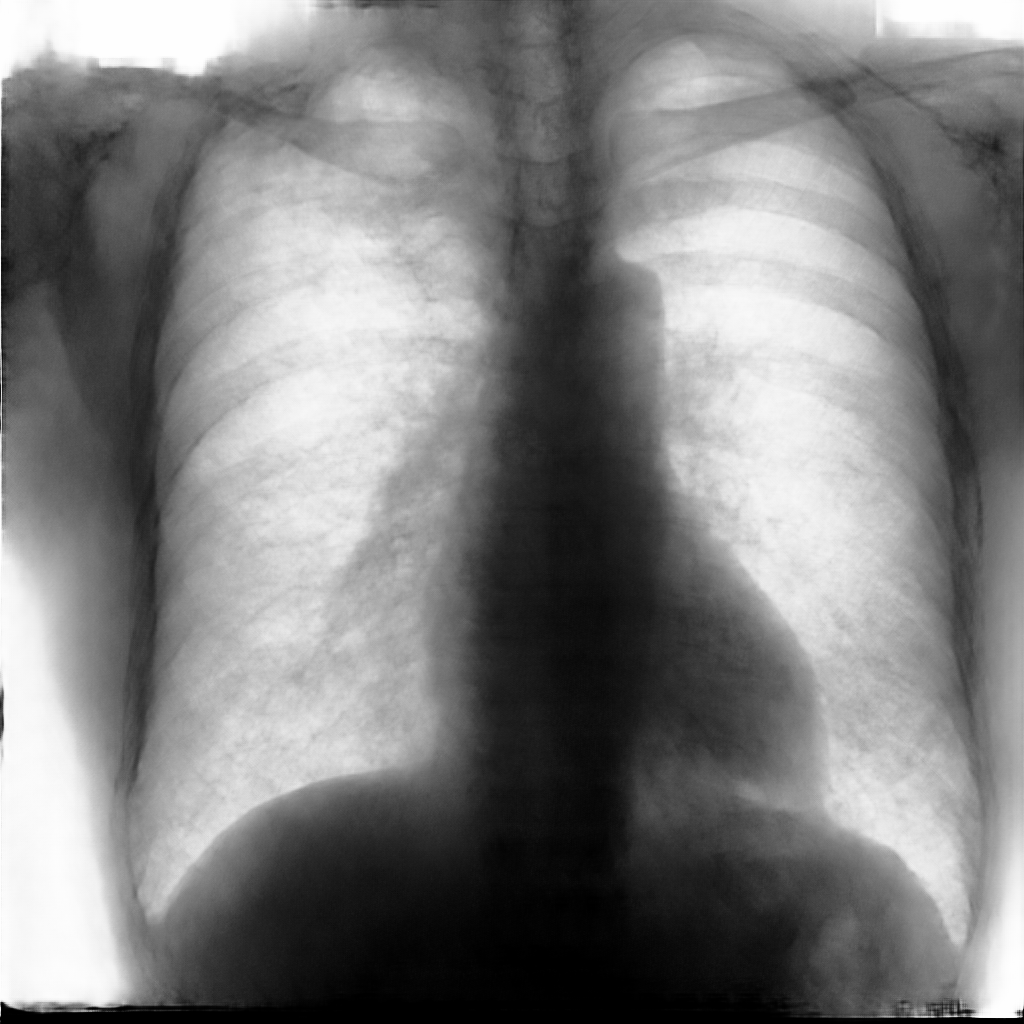} 
\includegraphics[width=2.63cm,height=2.63cm]{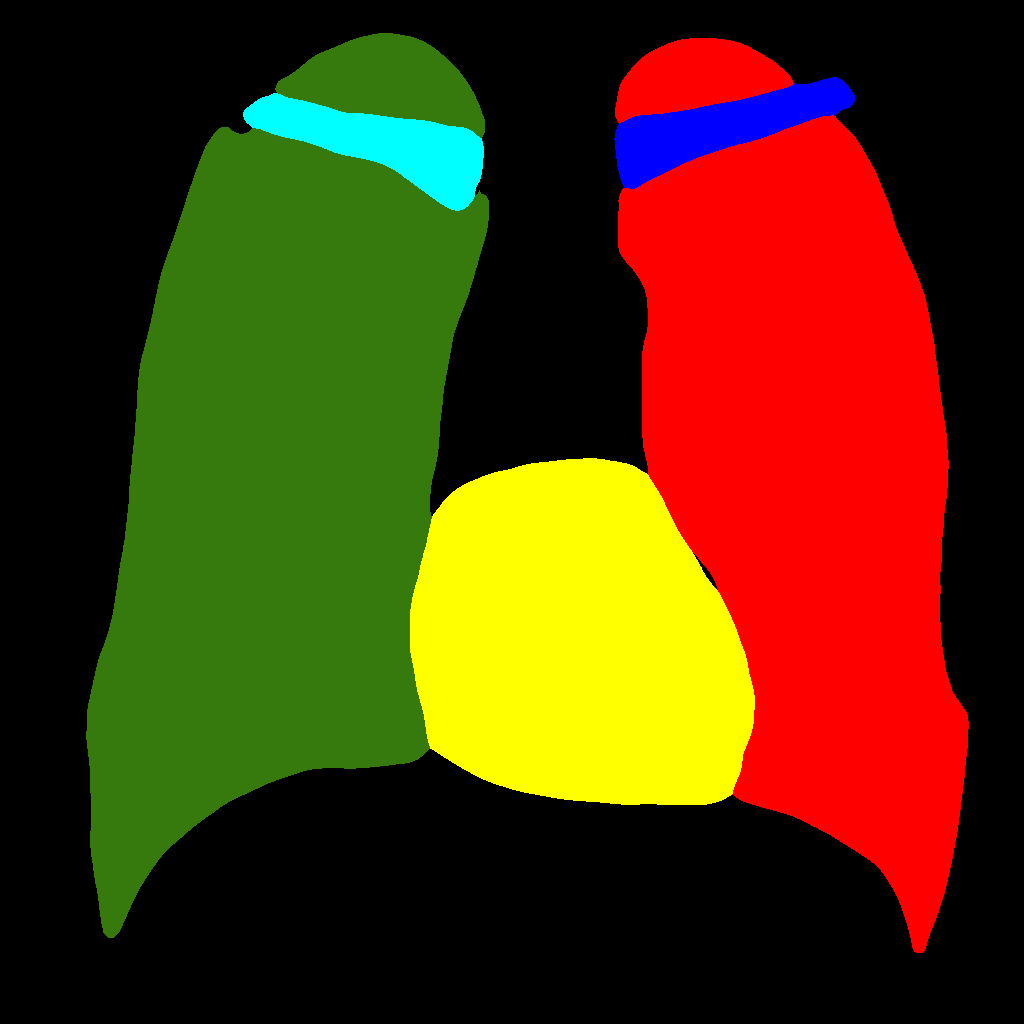}
\includegraphics[width=2.63cm,height=2.63cm]{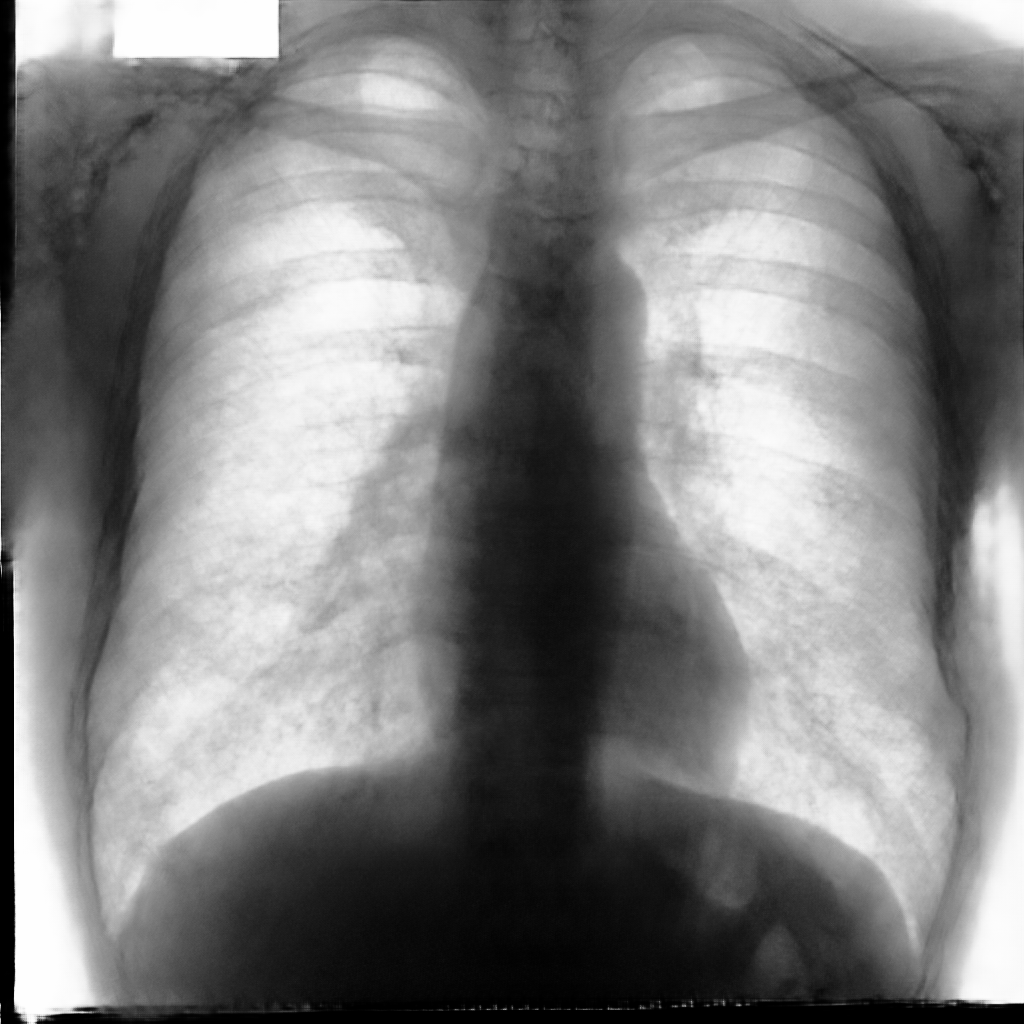}
\includegraphics[width=2.63cm,height=2.63cm]{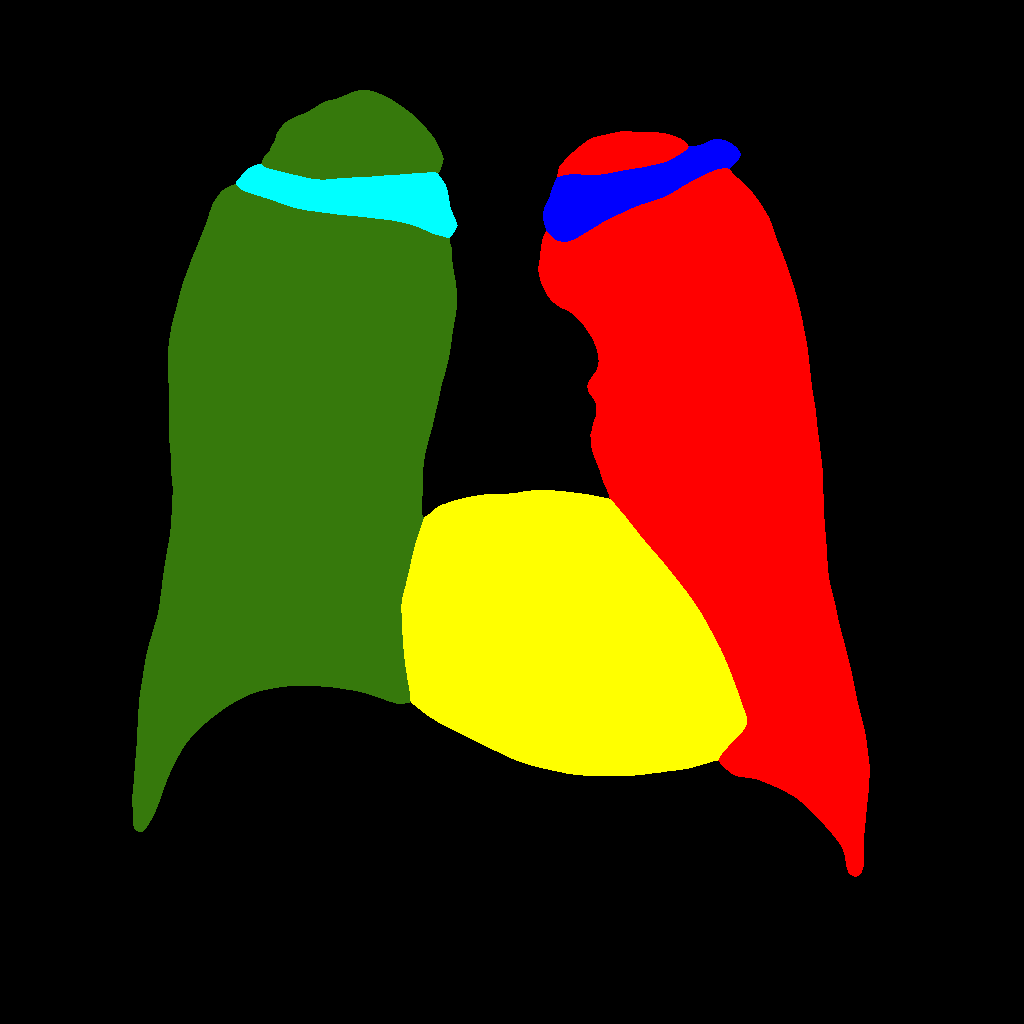}
\includegraphics[width=2.63cm,height=2.63cm]{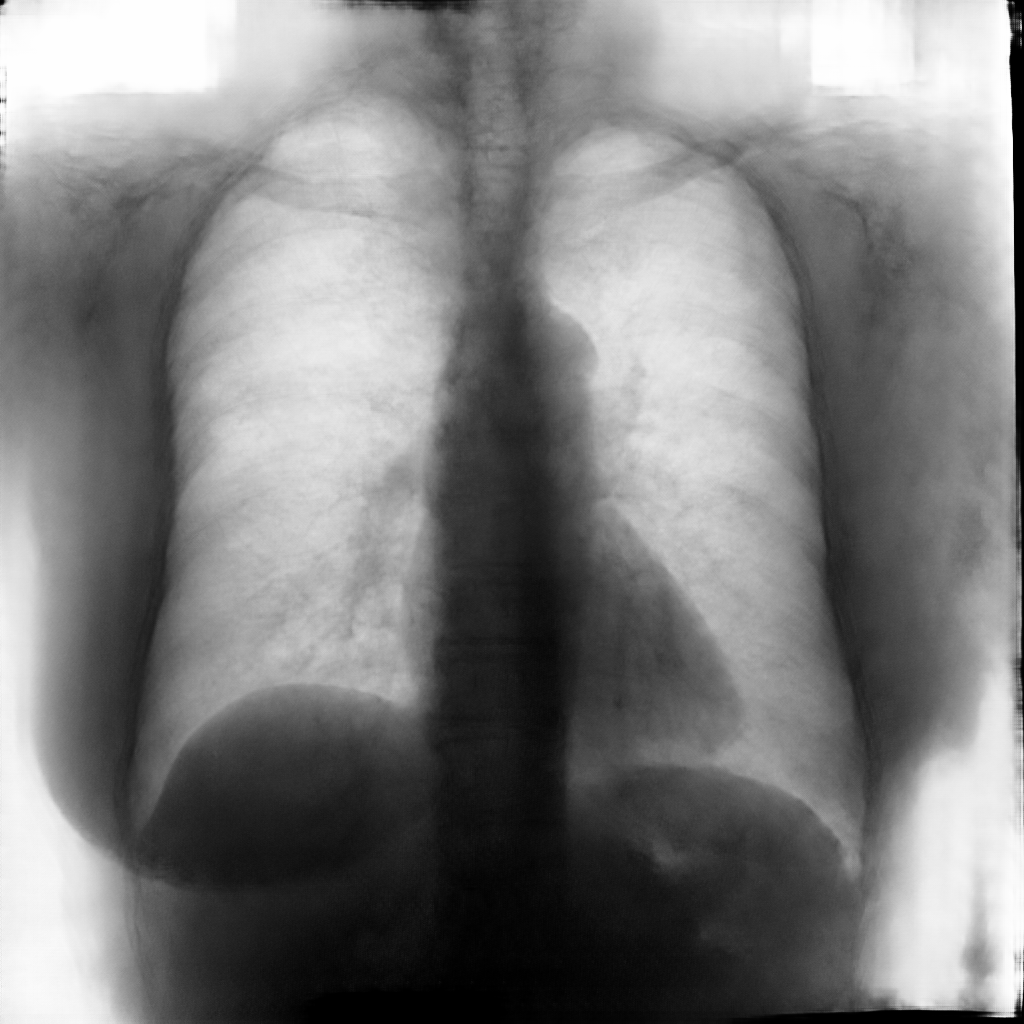}
} 
\end{center}
\caption{Examples of generated images based on the TINY\_DATASET.}
\label{fig:gen_examples_tiny}
\end{figure*}

Figure \ref{fig:gen_examples_full} and Figure \ref{fig:gen_examples_tiny} display some examples --- randomly chosen from all the generated images --- of the label--maps and the corresponding chest X--ray images generated with the three methods described in Section \ref{generation}, using the FULL\_DATASET and the TINY\_DATASET, respectively. We can observe that, with the single and two--stage methods, the images tend to be more similar to those belonging to the training set. For example, in most of the generated images there are white rectangles, which resemble those present in the training images, used to cover the names of both the patient and the hospital. 
Instead, the three--stage method does not produce such artifacts, suggesting that it is less prone to overfitting.

\begin{figure*}[t!]

\begin{center}
\subfloat[NODULES001.]{
\includegraphics[width=4.5cm,height=4.5cm]{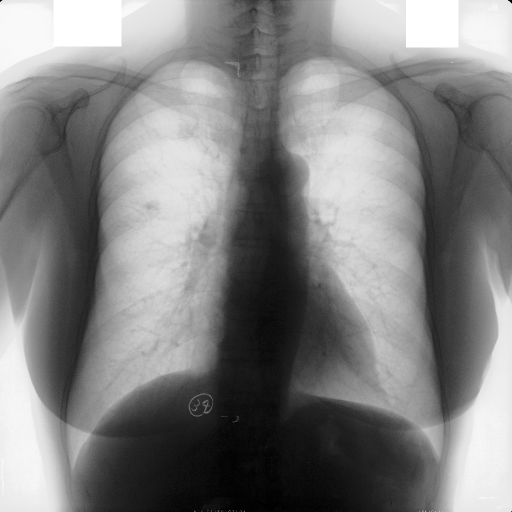}
\includegraphics[width=4.5cm,height=4.5cm]{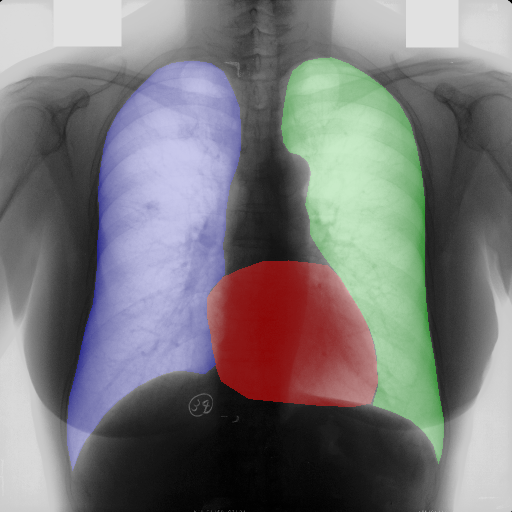}
\includegraphics[width=4.5cm,height=4.5cm]{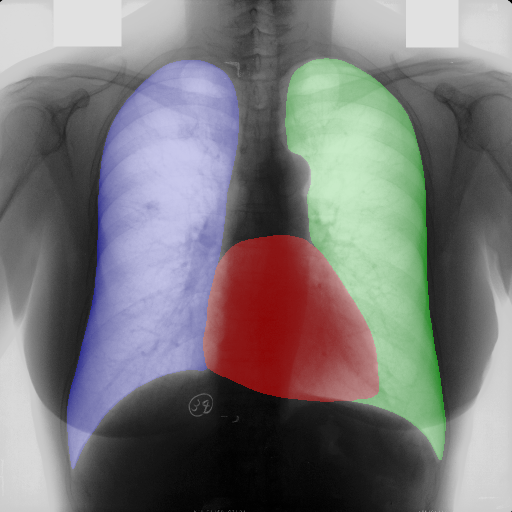} 
} \\
\subfloat[NODULES066.]{
\includegraphics[width=4.5cm,height=4.5cm]{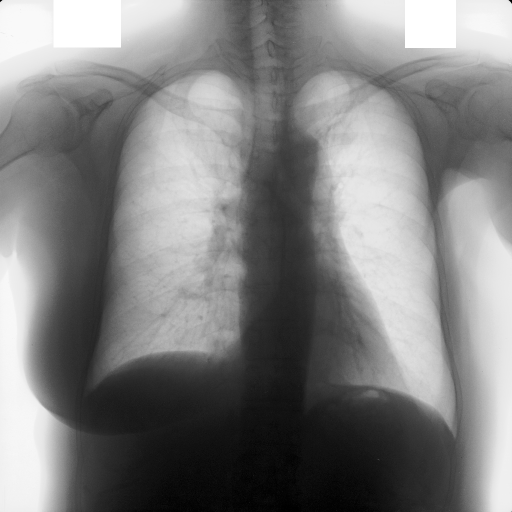}
\includegraphics[width=4.5cm,height=4.5cm]{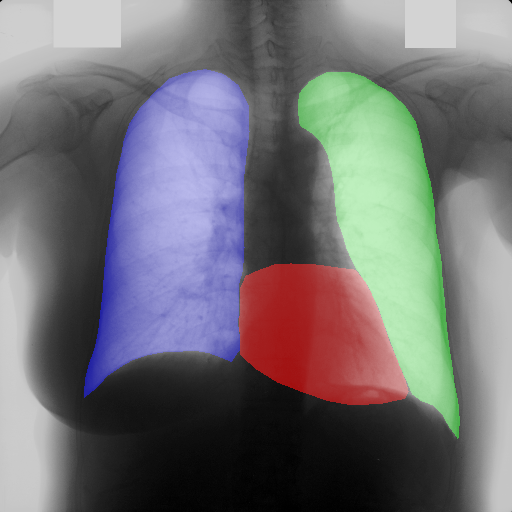}
\includegraphics[width=4.5cm,height=4.5cm]{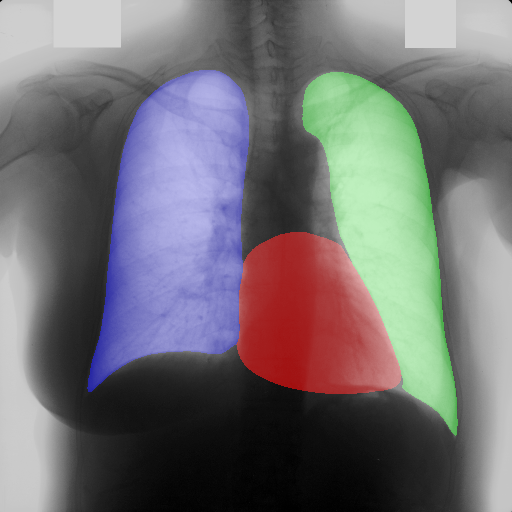} 
} 
\end{center}
\caption{Examples of segmented images for which doctors shared the same opinion. The first column represents the chest X-ray image, while the second and third columns are the target and our predicted segmentation, respectively.}
\label{fig:seg_examples_good}
\end{figure*}

\begin{figure*}[t!]

\begin{center}
\subfloat[NODULES014.]{
\includegraphics[width=4.5cm,height=4.5cm]{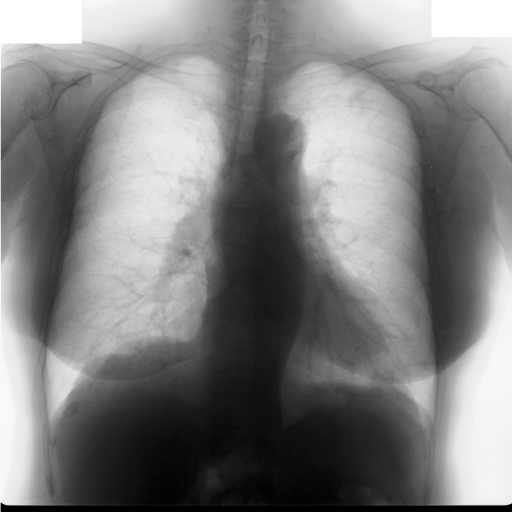}
\includegraphics[width=4.5cm,height=4.5cm]{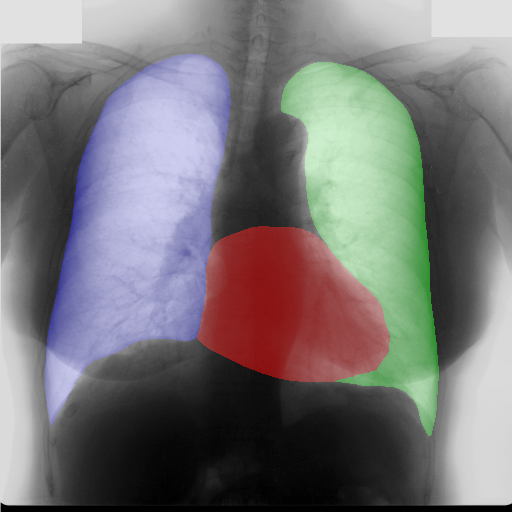}
\includegraphics[width=4.5cm,height=4.5cm]{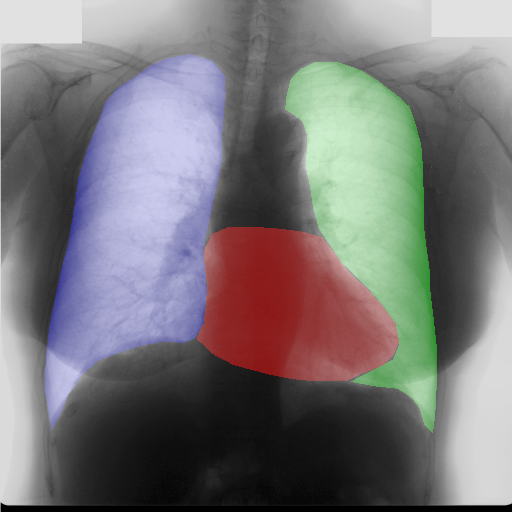} 
} \\
\subfloat[NODULES015.]{
\includegraphics[width=4.5cm,height=4.5cm]{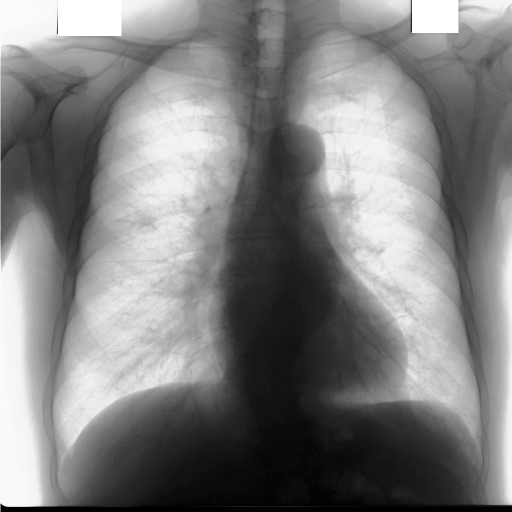}
\includegraphics[width=4.5cm,height=4.5cm]{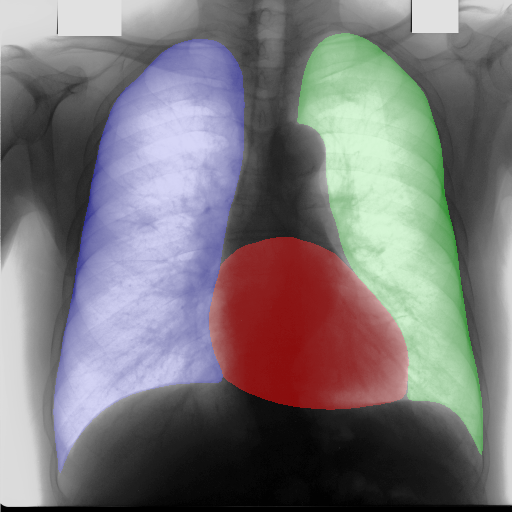}
\includegraphics[width=4.5cm,height=4.5cm]{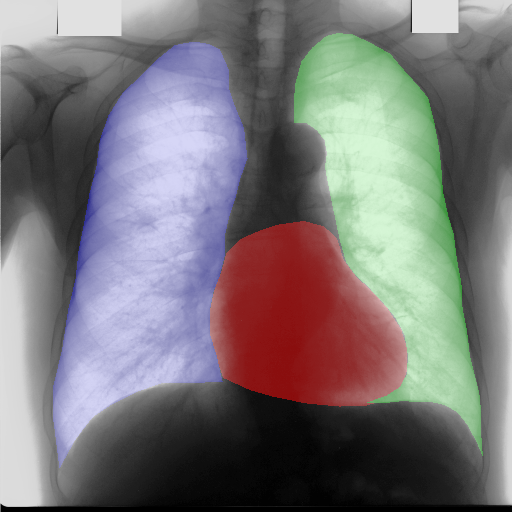} 
} 
\end{center}
\caption{Examples of segmented images for which doctors gave conflicting opinions. The first column represents the chest X--ray image, while the second and third columns are the target and our predicted segmentations, respectively.}
\label{fig:seg_examples_bad}
\end{figure*}

Moreover, in order to clarify the limits of the three--stage method, we assessed the quality of the segmentation results based on three human experts, who were asked to check 20 chest X--ray images, along with the corresponding supervision and the segmentation obtained by the SMANET network. Such images were chosen among those that can be considered difficult, at least based on the high error obtained by the segmentation algorithm. Figure \ref{fig:seg_examples_good} and Figure \ref{fig:seg_examples_bad} show different examples of the images evaluated by the experts. The first column represents the chest X--ray image, while the second and the third columns, whose order
was randomly exchanged during the presentation to the experts, represent the target segmentation and our prediction, respectively.
The three physicians were asked to choose the best segmentation and to comment about their choice.
Apart from a general agreement of all the doctors on the good quality of both the target segmentation and the segmentation provided by the three--stage method, surprisingly, they often chose the second one.
For the examples in Figure \ref{fig:seg_examples_good}, for instance, all the experts share the same opinion, preferring the segmentation obtained by the SMANET over the ground--truth segmentation. To report the results of the qualitative analysis, we numbered the target and predicted segmentation with 1 and 2, respectively, while doctors were assigned unordered pairs to obtain an unbiased result. Then, with respect to Figure \ref{fig:seg_examples_good}(a), the comments reported by the experts were: 1) In segmentation 1, a fairly large part of the upper left ventricle is missing; 2) I choose the segmentation number 2 because the heart profile does not protrude to the left of the spine profile; 3) The best is No. 2, the other leaves out a piece of the left free edge of the heart, in the cranial area. Instead, for Figure \ref{fig:seg_examples_good}(b), we obtained: 1) The second image is the best for the cardiac profile. For lung profiles, the second image is always better. The only flaw is that it leaks a bit on the right and left costophrenic sinuses. 2) Image 2 is the best, because the lower cardiac margin is lying down and does not protrude from the diaphragmatic dome. Image number 1 has a too flattened profile of the superior cardiac margin. 3) No. 2 for the cardiac profile more faithful to the real contours.

Instead, they reported conflicting opinions or decided not to give a preference with respect to the examples in Figure \ref{fig:seg_examples_bad}.
When they agreed, they generally found different reasons for choosing one segmentation over the other. With respect to Figure \ref{fig:seg_examples_bad}(a) the comments reported by the experts were: 1) I prefer not to indicate any options because the heart image is completely subverted; 2) Segmentation number 2 is better, even if it is complicated to read because there is a  ``bottle--shaped'' heart. The only thing that can be improved in image 2 is that a small portion of the right side of the heart is lost; 3) No. 1 respects more what could be the real contours of the heart image. Instead, for Figure \ref{fig:seg_examples_bad}(b) we obtained: 1) I prefer No. 2 because the tip of the heart is well placed on the diaphragm and does not let us see that small wedge--shaped image that incorrectly insinuates itself between heart and diaphragm in image 1 and which has no correspondence in the RX; 2) Both are good segmentations. Both have small problems, for example: in segmentation 1 a small portion of the tip (bottom right of the image) of the heart is missing, in segmentation 2 a part of the outflow cone (the ``upper'' part of the heart) is missing. It is difficult to choose, probably better No. 1 because of the heart; 3) No. 2 because No. 1 carnally probably exceeds the real dimensions of the cardiac image, including part of the other mediastinal structures.

These different evaluations, albeit limited by the small number of examined images, confirm the difficulty of  segmenting CXRs, a difficulty that is likely to be more evident in the case of the images selected for our quality analysis, which were chosen based on the large error produced by the segmentation algorithm.

\section{Conclusions}   \label{conclusions}
In this paper, we have proposed a multi--stage method based on GANs to generate multi--organ segmentation of chest X--ray images. Unlike existing image generation algorithms, in the proposed approach, generation occurs in three stages, starting with ``dots'', which represent anatomical parts, and initially involves low--resolution images. After the first step, the resolution is increased to translate ``dots'' into label--maps. We performed this step with Pix2PixHD, thus making the information grow and obtaining the labels for each anatomical part taken into consideration. Finally, Pix2PixHD is also used for translating the label--maps into the corresponding chest X--ray images. The usefulness of our method was demonstrated especially when there were few images in the training set, an affordable problem thanks to the multi--stage nature of the approach. 

It is worth observing that our method can be employed for any type of images, not exclusively medical ones, while synthetic and real images can concur in solving the segmentation problem (being used for pre--training and for fine--tuning the segmentation network, respectively), with a significant increase in performance. As a matter of future research, the proposed approach will be extended to other, more complex domains, such as that of natural images.  

\section*{Acknowledgements}
In addition to Dr. Tommaso Mazzierli, who is one of the authors of this work, we would like to thank Dr. Gabriella Gaudino and Dr. Valentina Vellucci for their contribution in the analysis of the segmentations.

\bibliographystyle{unsrt}  
\bibliography{references}  %%% Remove comment to use the external .bib file (using bibtex).

\end{document}